\documentclass[apj,twocolumn]{emulateapj}

\shorttitle{Profiles of Lyman $\alpha$ Emission Lines}
\shortauthors{Yamada et al.}

\newfont{\Sc}{eusm10}

\begin{document}

\title{Profiles of Lyman $\alpha$ Emission Lines of the Emitters at $z=3.1$}

\author{T. Yamada, \altaffilmark{1}, Y. Matsuda\altaffilmark{2}, K. Kousai\altaffilmark{3}, T. Hayashino\altaffilmark{3}, N. Morimoto\altaffilmark{1},  M. Umemura\altaffilmark{4} }
\email{yamada@astr.tohoku.ac.jp}

\altaffiltext{1}
{Astronomical Institute, Tohoku University, Aramaki,
Aoba-ku, Sendai, Miyagi, 980-8578}
\altaffiltext{2}
{Department of Physics, Durham University, South Road, Durham DH1 3LE}
\altaffiltext{3}
{Research Center for Neutrino Science, Graduate School of Science, Tohoku Univsersity, Sendai 980-8578, Japan}
\altaffiltext{4}
{Center for Computational Physics, University of Tsukuba, Tsukuba, Ibaraki 305, Japan}

\begin{abstract}

 We present the results of the observations of the Ly$\alpha$ line profiles of 91 emission-line galaxies at $z=3.1$ with the spectral resolution of $\lambda$/$\delta\lambda$(FWHM) $\approx 1700$, or 180 km s$^{-1}$. A significant fraction, $\sim 50$$\%$ of the observed objects show the characteristic double peaks in their Ly$\alpha$ profile. The red peak is much stronger than the blue one for most of the cases. The red peaks themselves also show weak but significant asymmetry and their widths are correlated with the velocity separation of the red and the blue peaks, which implies that the peaks are not isolated multiple components with different velocities but the parts of the single line which is modified by the absorption and/or scattering by the associated neutral hydrogen gas. The characteristic profile can be naturally explained by the scattering in the expanding shell of neutral hydrogen surrounding the Ly$\alpha$ emitting region while the attenuation by the inter-galactic medium should also be considered. Our results suggest that the star-formation in these Ly$\alpha$ emitters are dominated by the young burst-like events which produce the intrinsic Ly$\alpha$ emission as well as the gas outflow.

\end{abstract}

\keywords{galaxies: formation --- galaxies: evolution --- galaxies: emission}
\section{Introduction}

%%%%%%%%%%%%%%%%%%%%%%%%%%%%%%%%%%%%%%%%
\begin{figure*}[t!]
\includegraphics[width=17cm]{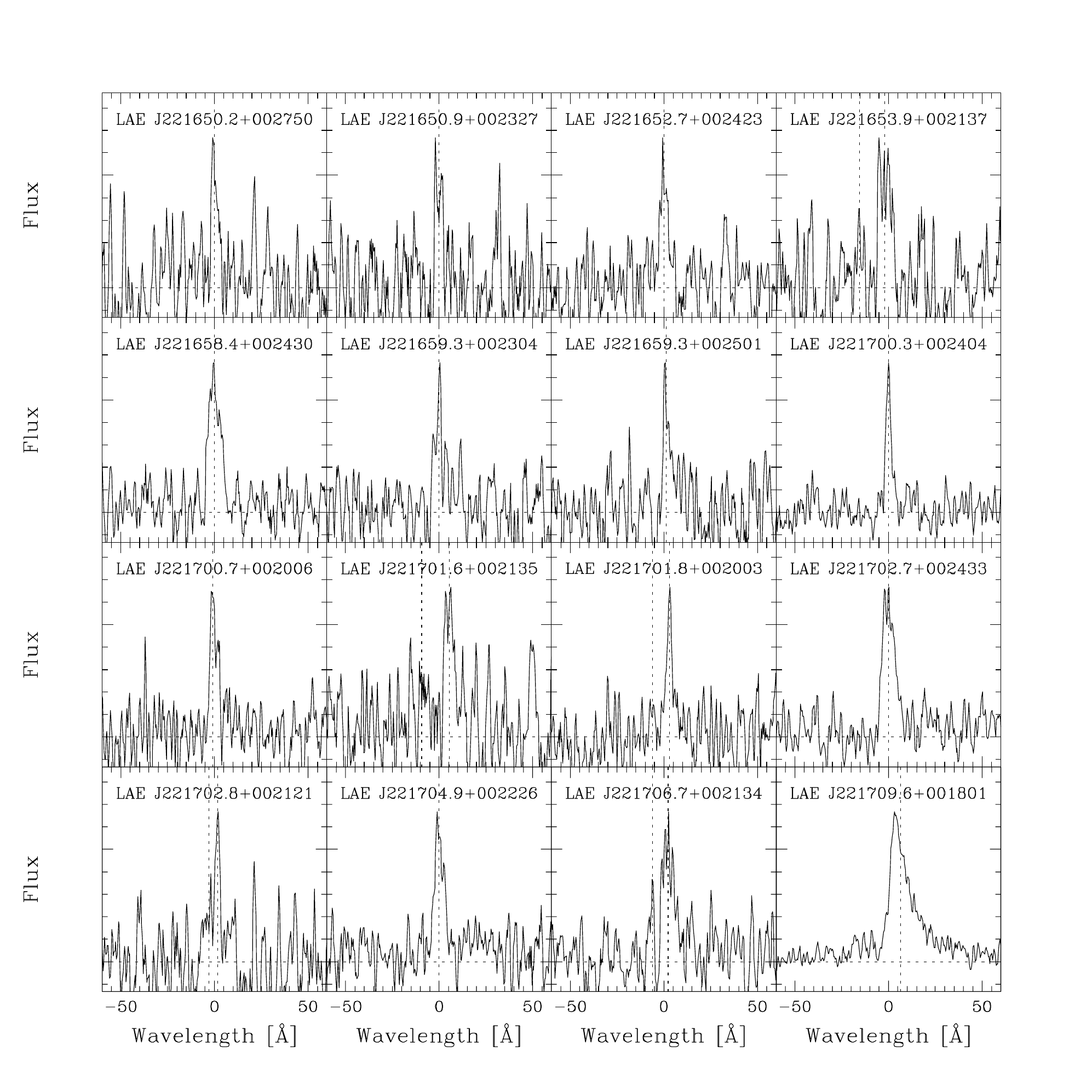}
\caption{The Ly$\alpha$ profile of the 91 $NB497$-selected emission-line objects at $z=3.1$ in SSA22-Sb1. The plot is centered at the central wavelength of the single Gaussian fitting. The vertical dotted lines show the center of the main and the significant secondary peaks in the multi-component Gaussian fitting.}
\label{fig1a}
\end{figure*}
%%%%%%%%%%%%%%%%%%%%%%%%%%%%%%%%%%%%%%%%
\figurenum{1}
%%%%%%%%%%%%%%%%%%%%%%%%%%%%%%%%%%%%%%%%
\begin{figure*}[t!]
\includegraphics[width=17cm]{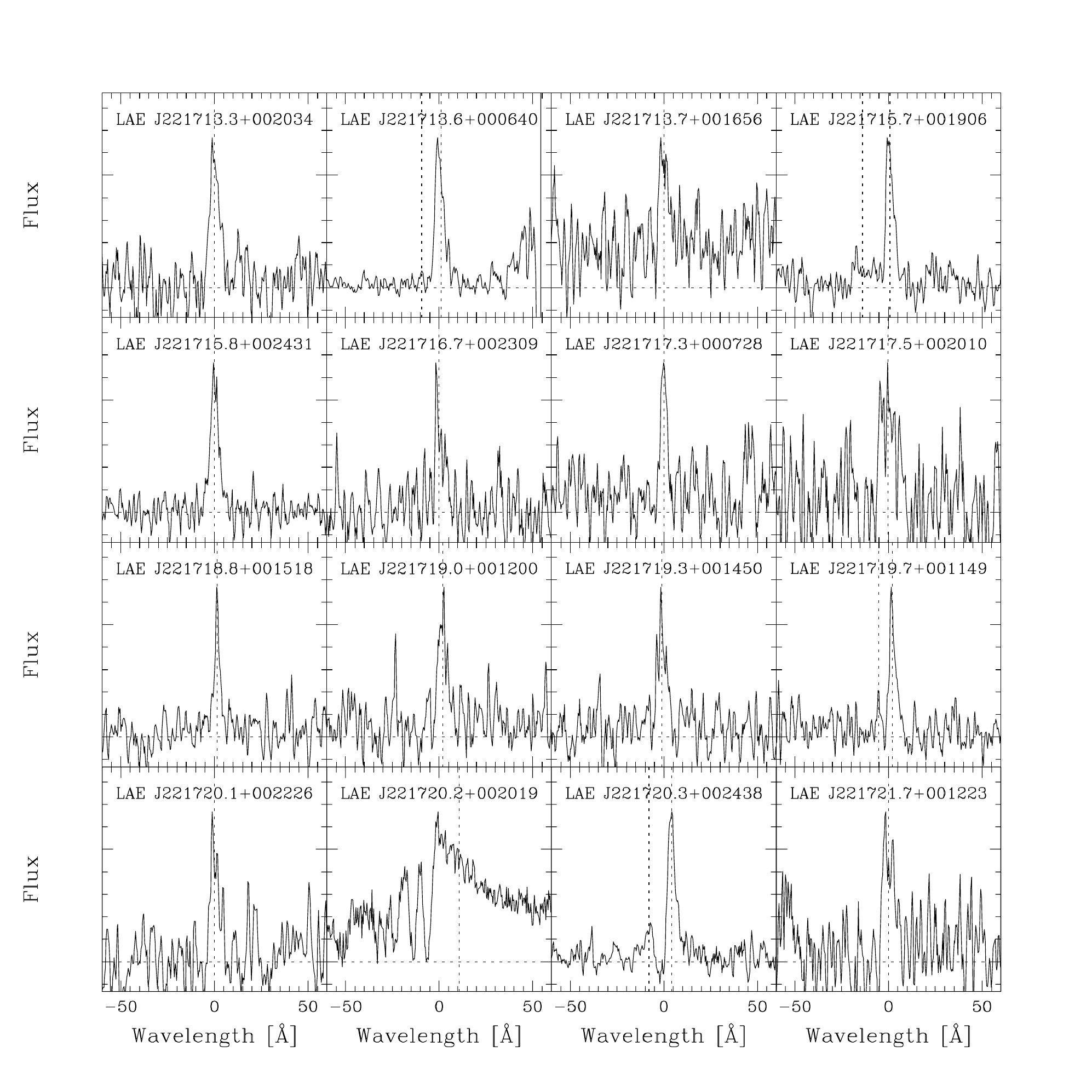}
\caption{Continued}
\label{fig1b}
\end{figure*}
%%%%%%%%%%%%%%%%%%%%%%%%%%%%%%%%%%%%%%%%
\figurenum{1}
%%%%%%%%%%%%%%%%%%%%%%%%%%%%%%%%%%%%%%%%
\begin{figure*}[t!]
\includegraphics[width=17cm]{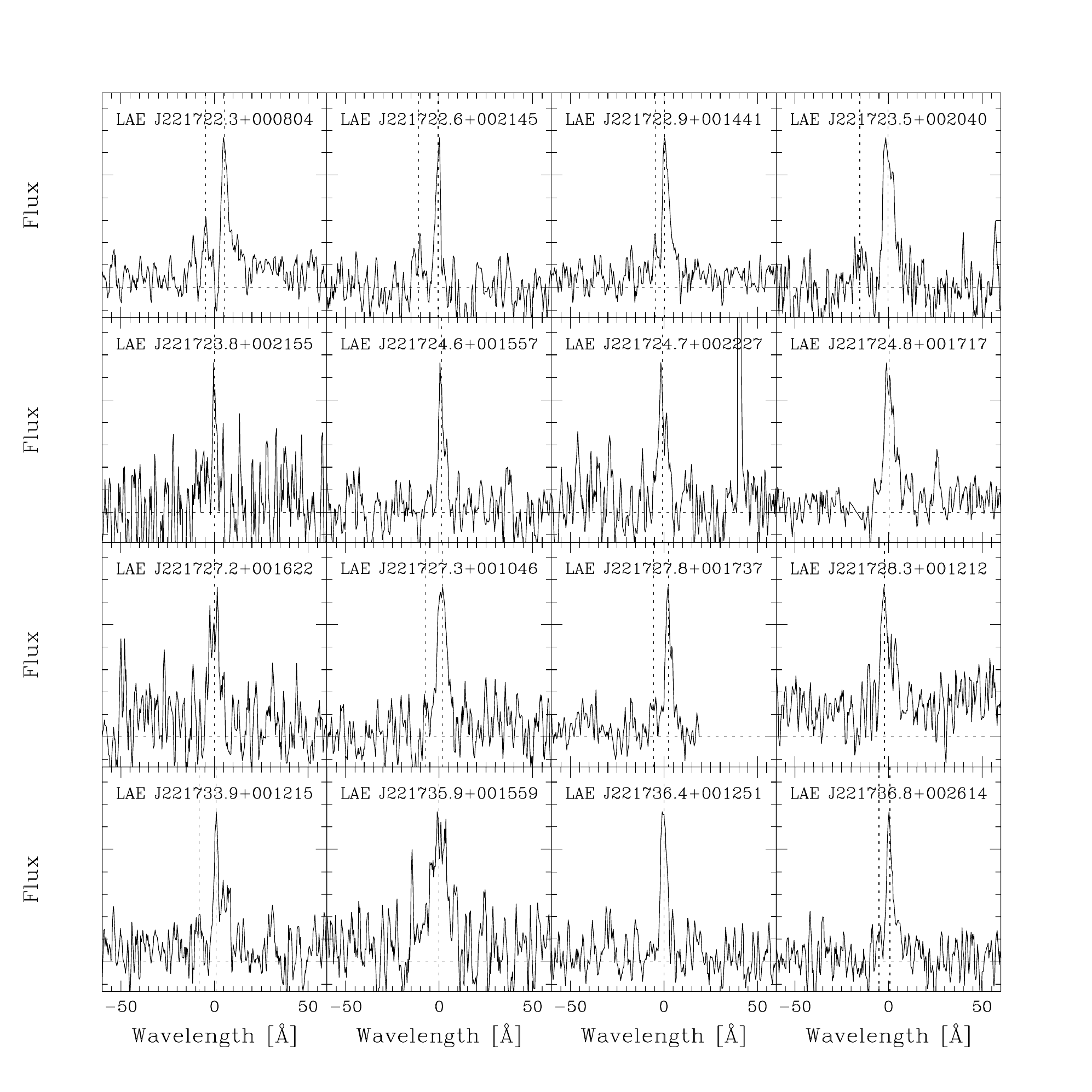}
\caption{Continued}
\label{fig1c}
\end{figure*}
%%%%%%%%%%%%%%%%%%%%%%%%%%%%%%%%%%%%%%%%
\figurenum{1}
%%%%%%%%%%%%%%%%%%%%%%%%%%%%%%%%%%%%%%%%
\begin{figure*}[t!]
\includegraphics[width=17cm]{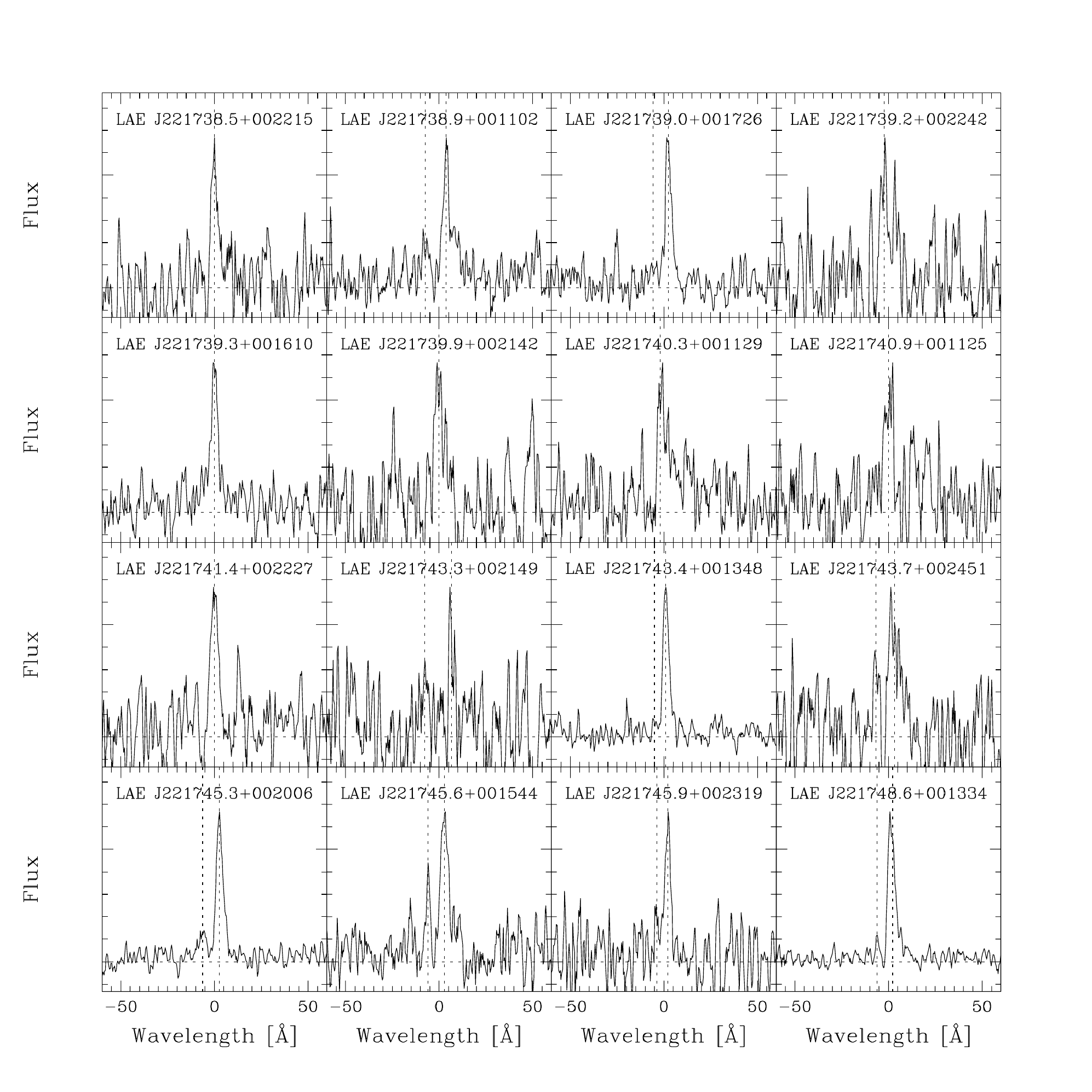}
\caption{Continued}
\label{fig1d}
\end{figure*}
%%%%%%%%%%%%%%%%%%%%%%%%%%%%%%%%%%%%%%%%
\figurenum{1}
%%%%%%%%%%%%%%%%%%%%%%%%%%%%%%%%%%%%%%%%
\begin{figure*}[t!]
\includegraphics[width=17cm]{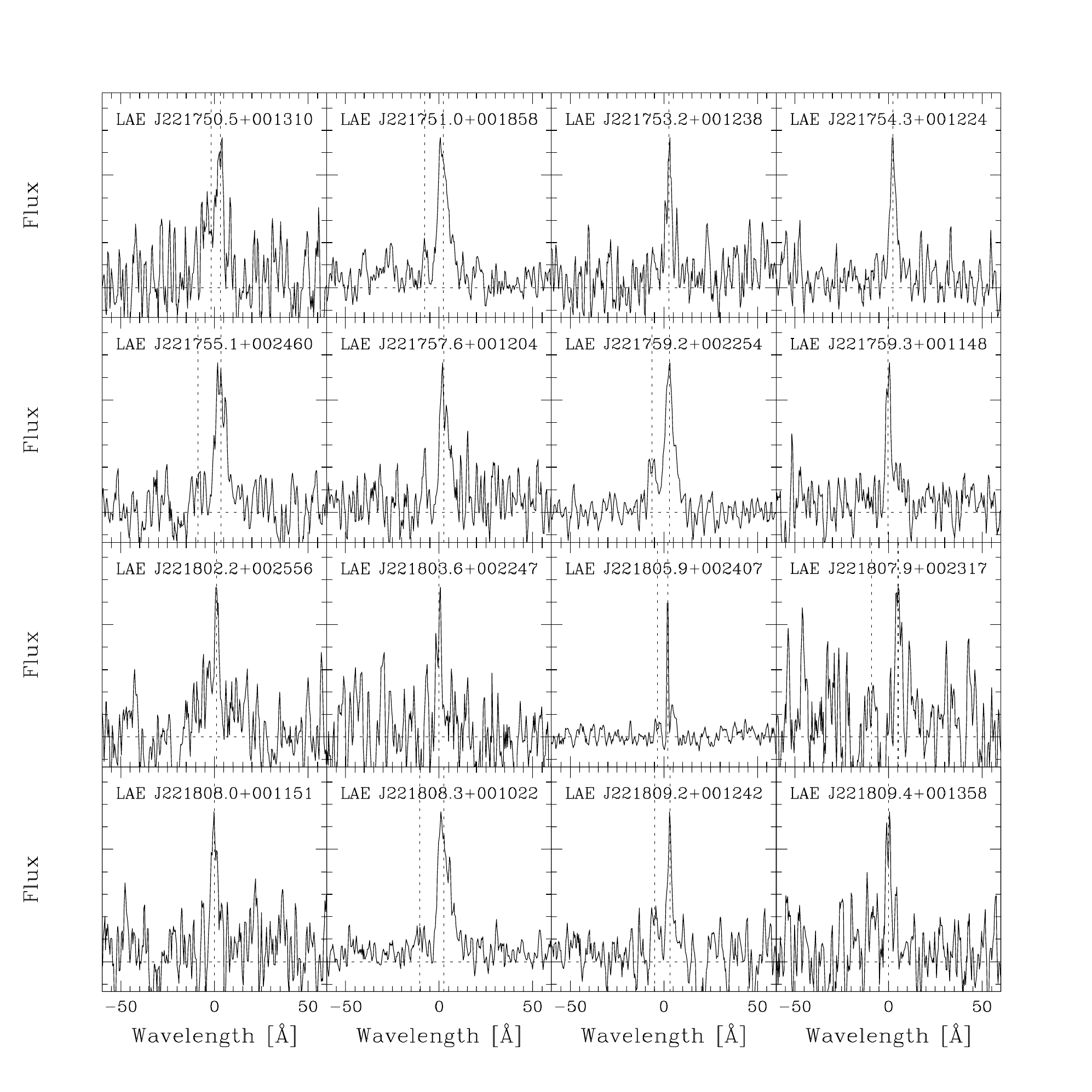}
\caption{Continued}
\label{fig1e}
\end{figure*}
%%%%%%%%%%%%%%%%%%%%%%%%%%%%%%%%%%%%%%%%
\figurenum{1}
%%%%%%%%%%%%%%%%%%%%%%%%%%%%%%%%%%%%%%%%
\begin{figure*}[t!]
\includegraphics[width=17cm]{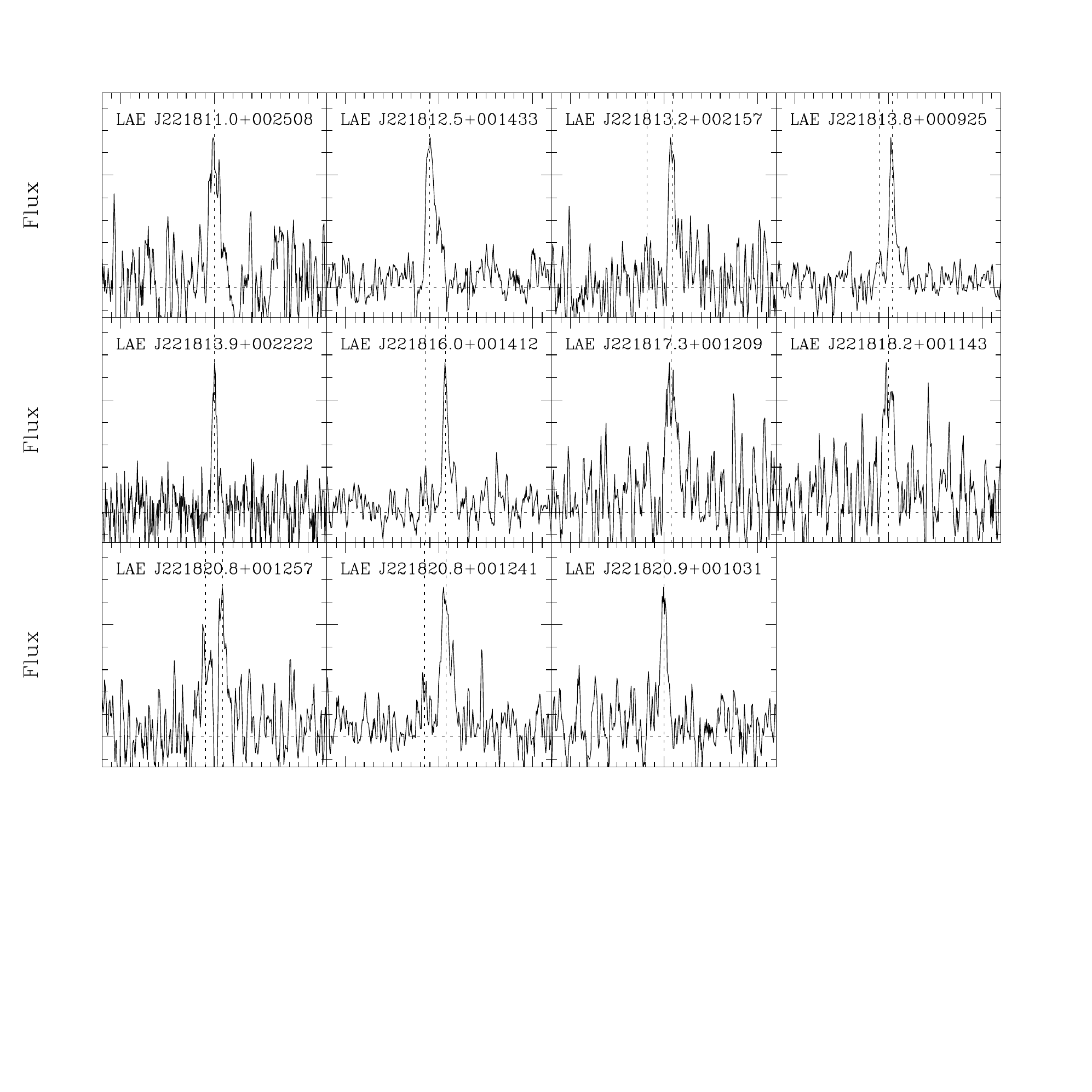}
\caption{Continued}
\label{fig1f}
\end{figure*}
%%%%%%%%%%%%%%%%%%%%%%%%%%%%%%%%%%%%%%%%

 A large number of the Ly$\alpha$ emitters at high redshift have been detected in sensitive search using the narrow-band filters (e.g., Hu \& McMahon 1996; Steidel et al. 2000; Rhoads et al. 2001; 2003; Kodaira et al. 2003; Palunas et al. 2004; Hayashino et al. 2004; Matsuda et al. 2004; Taniguchi et al. 2005; Gronwall et al. 2006; Iye et al. 2006, Ouchi et al. 2008). From the photometric study, they are generally considered to be young star-forming objects and found to be widely distributed in their luminosity, size, and equivalent width (e.g., Gawiser et al. 2007; Finkelstein et al. 2007).

 While the Ly$\alpha$ emission is one of the fundamental tools to study the galaxy formation at high redshift, it is, however, still difficult to understand what the dominant origins of the Ly$\alpha$ emission are, and how the Ly$\alpha$ photons escape from the galaxy. Resonance scattering and extinction by the neutral hydrogen or dust in galaxies may easily modify the observed flux and line profile. There are at least three different physical origins of the Ly$\alpha$ lines from high-redshift star-forming galaxies; photo-ionization by hot massive stars, cooling radiation from the gas heated either by the shock during to the gravitational collapse or by the shock due to the gas outflow driven by the thermal energy of frequent supernovae or active galactic nuclei. The Ly$\alpha$ photons produced by any of these process are affected by the neutral hydrogen gas mixed with or surrounding the emission-line regions. Recent results suggest that statistical average of the Ly$\alpha$ escape fraction of a sample of the star-forming galaxies at redshift $z \sim 2$ is only $\sim 5$$\%$ (Hayes et al. 2009).

 It is important to study their spectroscopic properties, especially their line profiles to study how the lines are produced and modified. If the Ly$\alpha$ emission is largely reprocessed by the neutral gas with large bulk motion, we may observe the signatures in their line profiles. For example, Dijikstra et al. (2006) studied the expected Ly$\alpha$ profile of optically thick, spherically symmetric collapsing gas clouds. The expected profiles are generally blueshifted and show double peaked profile possibly dominated by the blue peak due to the combination of the infalling gas motion and line radiation transfer effects. On the other hand, Verhamme et al. (2006, 2008) modeled the line profiles expected for the expanding neutral hydrogen shell. Although the expected profiles are also double-peaked, in many cases they are redshifted and the red peak appears stronger due to the scattered components behind the original Ly$\alpha$ emission-line region in our line of sight. It is, however, not easy to discriminate these two cases from the observation of the individual Ly$\alpha$ line itself if we do not know the systemic velocity.  

 It is known that high-redshift star-forming galaxies typically show gas outflow from the sources. Pettini et al. (2001) and Shapley et al. (2003) compared the redshifts of the interstellar absorption line with the rest-frame optical emission lines which are nearly at the systemic velocity of the photo-ionized region and found that the inter-stellar absorption gas is generally blueshifted for typically 200 km s$^{-1}$ indicating that the outflow gas motion is common among the Lyman Break Galaxies. Ly$\alpha$ lines in these objects also show P Cyg-type profiles, namely, a narrow emission line with a blueshifted absorption line. Verhamme et al. (2008) applied their models to the Ly$\alpha$ line profile of the Lyman Break Galaxies and found that they can be explained by the scattering in the simple expanding shell models. Steidel et al. (2010) also showed that the cool gas around $z=2$-3 Lyman Break galaxies typically exhibit outward motion.
 
  On the other hand, the models of cold accretion (Dekel et al. 2009; Goerdt et al. 2009) also suggest that the observed luminosity, surface brightness, and size distributions of the Ly$\alpha$ emitters, especially for the extended Ly$\alpha$ Blobs (Steidel et al. 2000; Matsuda et al. 2004) can be explained by the gas heated by the collisional excitation in the cold gas stream.

 Spectroscopic observations to study the line profiles of the Ly$\alpha$ emitters are essential to discriminate these cases although only a small sample of spectroscopy with enough high spectral resolution is available so far.  Matsuda et al. (2006) studied the line profiles of the extended Ly$\alpha$ emitters, or Ly$\alpha$ Blobs as well as more 'ordinary' Ly$\alpha$ emitters, and found that the Ly$\alpha$ lines of the blobs have more structure and the whole velocity width is positively correlated with their size. 

 In this paper, we present the results of our new spectroscopic observations for the 91 Ly$\alpha$ emitters to show their Ly$\alpha$ line profiles. The photometric data for the sample selection is similar as in Matsuda et al. (2006) but the number of the spectra for normal Ly$\alpha$ emitters are significantly increased. We describe our observations and data reduction in Section 2, the observed line profiles in Section 3, and the discussion in Section 4.

\section{Observations and Data Reduction}

 The sample of the Ly$\alpha$ emitters are selected based on the same data presented in Hayashino et al. (2004) at the field of SSA22-Sb1 where one of the largest overdensity of Lyman Break Galaxies as well as Ly$\alpha$ emitters at $z$=3.1 is known to exist (Steidel et al. 1998; 2000). The selection criteria for the emitters are, however, slightly modified to include the objects with relatively low equivalent width. We also did not use the $B-V$ color criteria, which was needed to make the very robust imaging sample of Ly$\alpha$ emitters to exclude the contamination by the foreground objects. We here applied only the following criteria.

$$ BV-NB497 > 0.5 \qquad \& \qquad NB497 < 25.5 ,$$ or

$$ BV-NB497 > 1.0 \qquad \& \qquad NB497 < 26.0 ,$$

where $BV$ is the effectively averaged $AB$ magnitude of the $B$ and the $V$-band and $NB497$ represents the $AB$ magnitude in the NB497 narrow-band filter (Hayashino et al. 2004).

 The observations of the Ly$\alpha$ profiles of the emitters are done by the Faint Object Camera And Spectrograph (FOCAS) equipped with the Subaru 8.2m telescope in Jul 2004- Aug 2005. The VPH grating 600\_450 was used to obtain the spectral dispersion of 0.37 \AA\ per pixel. Using slits with a width of 1 arcsec and spectral resolution, $\lambda$/$\delta\lambda$ $\approx 1700$ was achieved as measured from the widths of the arc lines. In order to observe as many targets as we can in one exposure, we used the custom-made intermediate band filter with $\delta\lambda$ $\sim 200$\AA\ to limit the spectral length on the detectors. By this method, we obtain the range of the spectra only around the emission lines, but the spectral resolution is high enough to clearly discriminate the [OII] 3727 \AA\ doublet, which is the only significant contaminants, from the $z=3.1$ Ly$\alpha$ lines. The data was reduced in the standard manner using IRAF. The accuracy of the wavelength calibration is higher than 0.2 \AA\ . The one-dimensional spectra are then extracted by averaging the three spatial pixels, each 0.$"$2. We also smoothed the spectra by three pixels in the dispersion direction.

 As a result, we obtained the Ly$\alpha$ emission-line spectra for the 91 objects. The sample contains the 12 objects classified as Ly$\alpha$ Blobs by Matsuda et al. (2004) whose isophotal area above the surface brightness of $\approx 7 \times 10^{-17}$ erg s$^{-1}$ cm$^{-2}$ arcsec$^{-2}$ is larger than 16 arcsec$^2$ or 900 kpc$^2$ at $z=3.1$. Matsuda et al. (2004; 2006) showed that the photometric or the spectroscopic properties of the Ly$\alpha$ Blobs are not quite discriminated from other objects but rather continuously distributed in the whole sample of the emitters.

 As described in Matsuda et al. (2005) and Matsuda et al. (2006), the sample selected by the criteria in Hayashino et al. (2004) is almost free for contamination ($< 1\%$). For the objects with lower equivalent width (i.e., $0.5 < BV-NB497 < 1.2$), the fraction of the [OII] emitter at $z=0.33$ increases, as expected, but still less than $\sim 5$\% . We identified in total of 91 spectroscopic Ly$\alpha$ emitters in the 11 masks located in the SSA22-Sb1 field.

\section{Observed Line Profiles}

 Fig.1a-1f show the observed Ly$\alpha$ spectrum for all the 91 objects. Two of them (LAE J221709.6+001801 and LAE J221720.2+002019) clearly show a very large line width ($\gtrsim$ 1000 km s$^{-1}$) and possibly host active galactic nuclei (AGN). Table 1 summarizes their observed properties. 

 In Fig.1, we first notice that many of the objects show the characteristic profiles, namely the double-peaked spectra with a strong, asymmetrical red peak and a much weaker blue peak. While the most conspicuous cases are the objects LAE J221720.3+002438, LAE J221745.3+002006, and LAE J221759.2+002254 (LAB28 in Matsuda et al. 2004), many other cases can be recognized in Fig.1. As a large fraction of the observed Ly$\alpha$ spectra shows the similar profiles, it may represent the common physical properties of the emitters.

 We therefore tried to characterize the profile and to evaluate the fraction of the objects with the similar profile among those observed in more objective way. We first fitted the visually isolated peaks in each spectrum with the Gaussian profiles. These procedures are rather formal ones but useful in the following discussions. Fig.2 and Fig.3 show the example of the fitting for the two objects with high and moderate signal-to-noise ratio, respectively. Fig.2a and 3a are for the multiple Gaussian fitting. The green lines show the each component and the blue ones the total. The red dashed lines in the figures show the noise level measured at the same wavelength by using the blank part of the slits. Bottom panels show the residual. On the other hand, Fig.2b and 3b show the examples of the single Gaussian fitting similar with those tried in Matsuda et al. (2006). For the objects with multiple peaks, the data points between the peaks are not used for the fitting (thin crosses in Fig.2b and 3b). The center of the wavelength in each panel in Fig.1 is the central wavelength of the single-component Gaussian fitting. The central wavelengths of the primary red and secondary blue peaks (for the objects with the feature) as well as that of the single-component fitting (for all the objects) are also presented in Table 1.

\figurenum{2}
%%%%%%%%%%%%%%%%%%%%%%%%%%%%%%%%%%%%%%%%
\begin{figure}[h]
\includegraphics[width=8cm]{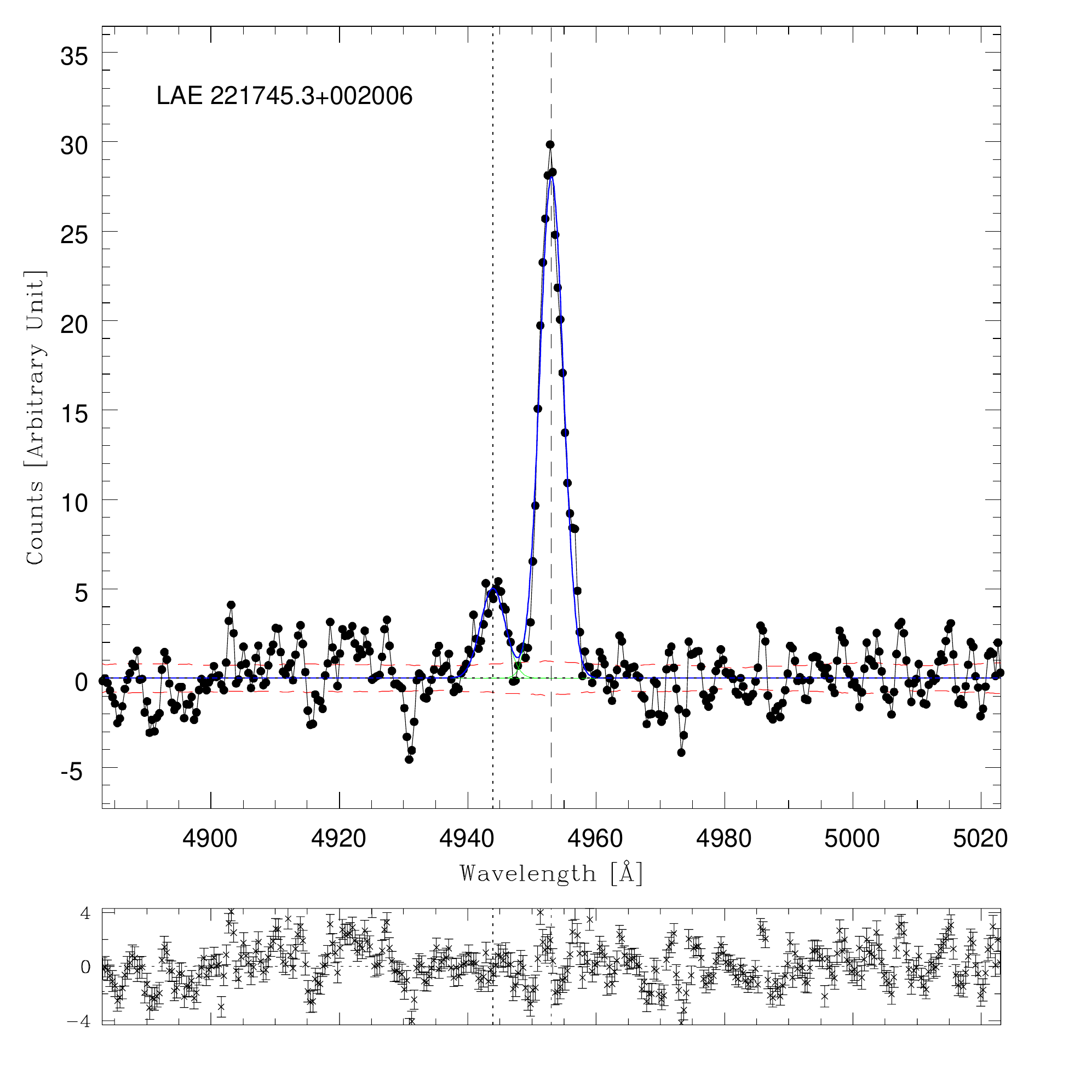}
\includegraphics[width=8cm]{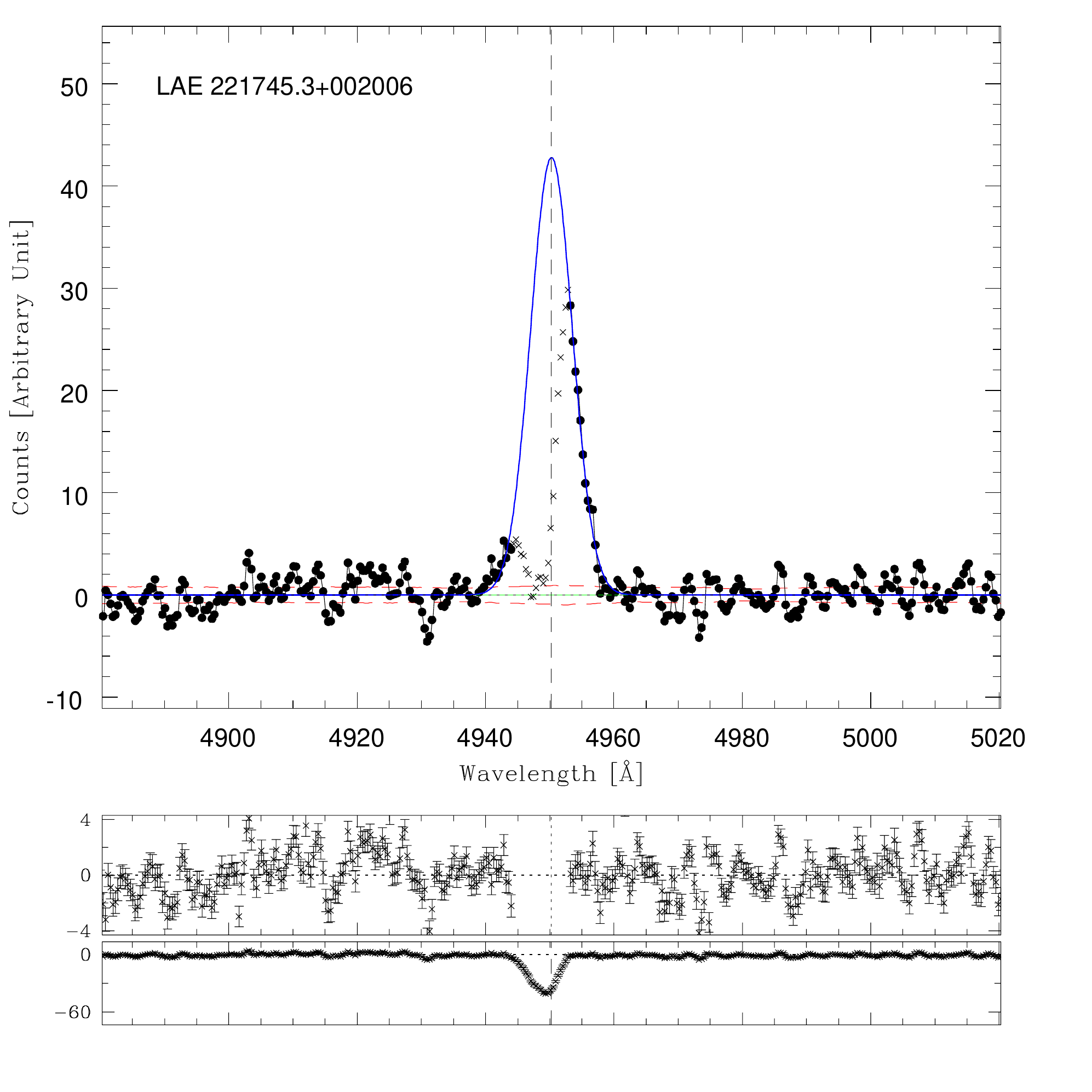}
\caption{(a) Example of the formal multiple Gaussian fitting to the profile of an object observed with the high signal-to-noise ratio. The blue line is the best-fitted Gaussian profiles and the red dashed lines show the background noise level (root mean square values).(b) Same as Fig.2a but for the formal single Gaussian fitting. The data points used in the fitting are plotted by the filled dots. Crosses are those not used in the fitting. The bottom panels show the formal residuals.
}
\label{fig1}
\end{figure}
%%%%%%%%%%%%%%%%%%%%%%%%%%%%%%%%%%%%%%%%

\figurenum{3}
%%%%%%%%%%%%%%%%%%%%%%%%%%%%%%%%%%%%%%%%
\begin{figure}[h]
\includegraphics[width=8cm]{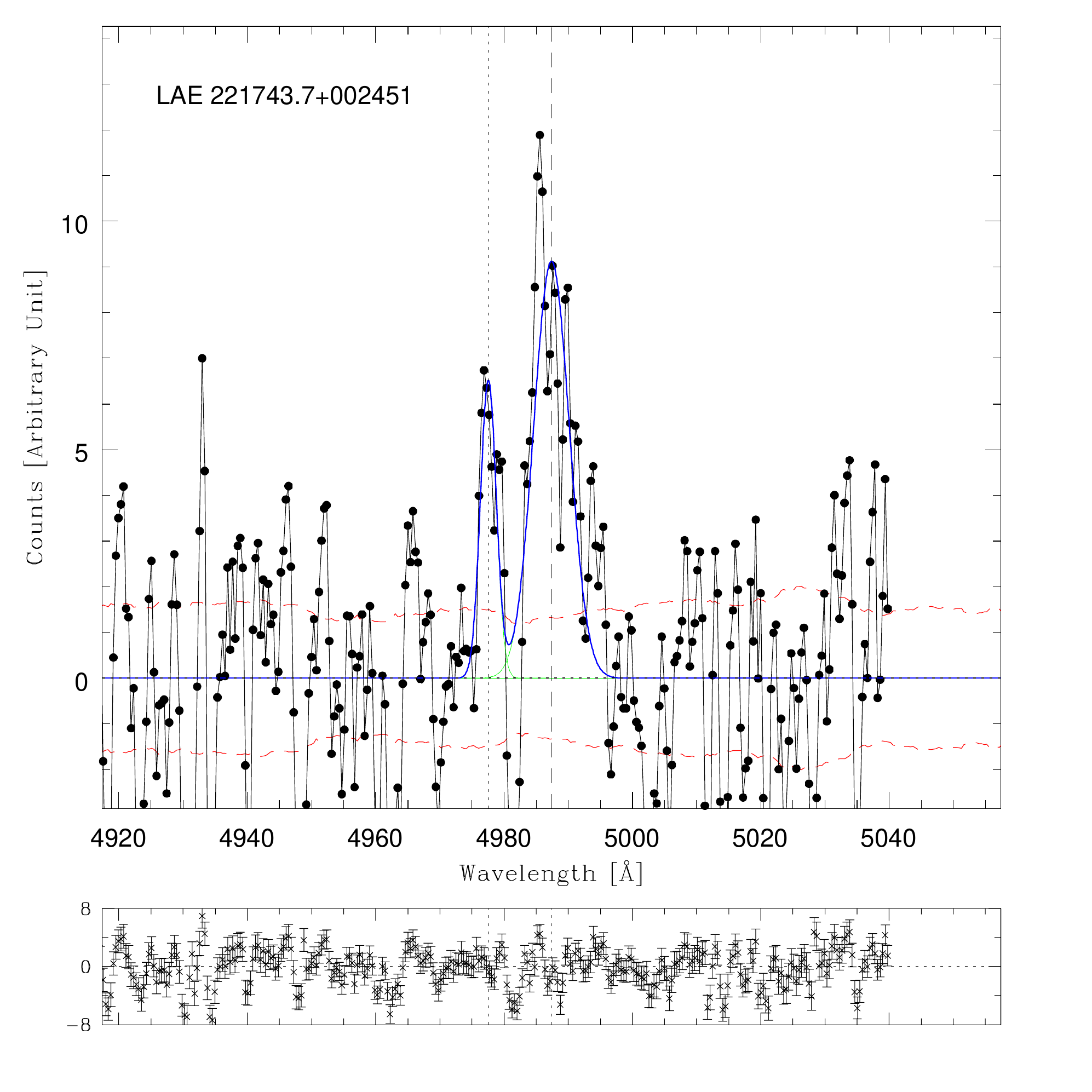}
\includegraphics[width=8cm]{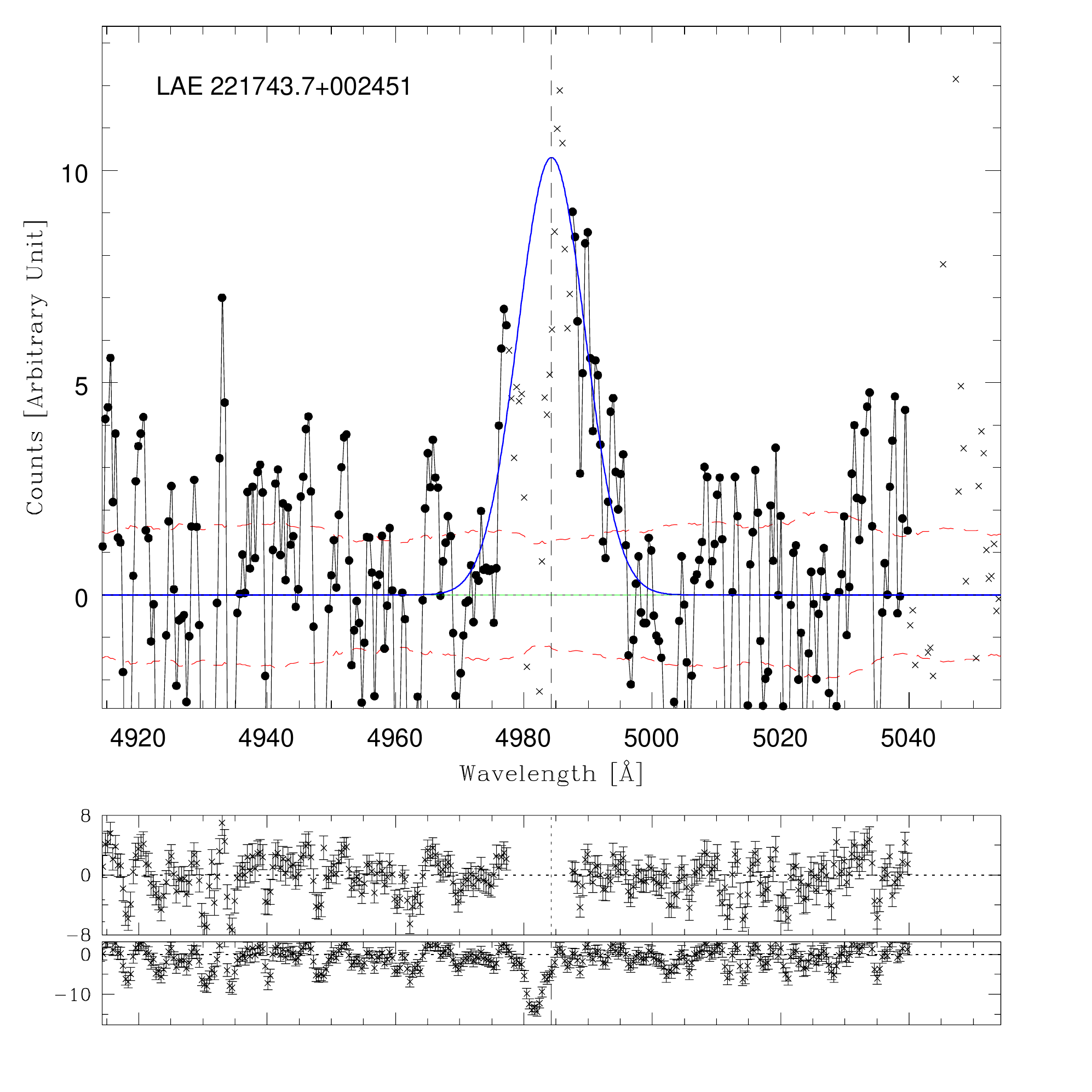}
\caption{(a) Same as Fig. 2a but for an object with the relatively poor S/N ratio.(b) Same as Fig. 2b but for an object with the relatively poor S/N ratio.
}
\label{fig3a}
\end{figure}
%%%%%%%%%%%%%%%%%%%%%%%%%%%%%%%%%%%%%%%%

 Of the 89 objects except for the two broad-line AGN, 50 objects have the visually identified strong blue and weak red peaks. 4 other objects (221728.3+001212, 221740.3+001129, 221740.9+001125 , and 221812.5+001433) do not show the weak blue peaks but have notable components to the red of the strongest peak. Three of these four objects, 221728.3+001212, and closely located objects 221740.3+001129 and 221740.9+001125 are associated with the Ly$\alpha$ Blobs LAB33 and LAB7 in Matsuda et al. (2004), respectively.  
 
 For these 50 candidates of the ``strong red and weak blue" profiles, we then evaluated the significance of the weak peaks relative to the noise level. We evaluated the r.m.s. noise in the blank-sky spectra (the red dashed lines in Fig.2 and Fig.3 for the example) at the line core ($\pm 2$$\sigma_\lambda$ of the Gaussian) and then selected only the objects with the weak peaks which are more significant than the 3$\sigma$ level as the final sample of the ``strong red and weak blue" profiles. For the cases between 3$\sigma$ and 4$\sigma$, we further visually checked the spectra and rejected a few cases. After all, 39 objects among the 89 (44$\%$) show the significant characteristic profile. The wavelength of the primary and the secondary peaks are shown by the vertical dotted lines in Fig.1.  This is a large fraction, which implies that the ``strong red and weak blue" profile is a common characteristic property of the observed Ly$\alpha$ emitters. The fraction should be considered even as the lower limit because it is  more difficult to detect the secondary weaker peaks significantly if the objects are faint or their emission-lines are weak. Indeed, if we limit the objects with $NB497 < 25$ and further with $BV-NB497 > 1.0$, the fraction is even larger, 53 and 55$\%$, respectively. Table 2 summarizes the number of the objects with the characteristic profile. We also listed the wavelength of the secondary component for five marginal objects whose blue peaks were visually identified but not found to be significant in Table 1 with ':'.

 While it is difficult, only from such formal fitting procedure as shown in Fig.2 and Fig.3, to tell whether the peaks are the multiple emission lines with different central velocities or the single line modified by the absorption and/or scattering by neutral gas in vicinity, the latter interpretation is plausible as such large fraction of the sample always show much stronger peak at the long wavelength. We confirm this further by the following two different methods.

 First, we investigated the asymmetry of the strongest peaks of the sample. We identify the pixel with the largest flux and obtained the ratio of the flux integrated over the 4.5\AA\ (about 1.5 times the spectral resolution) toward longer wavelength to those toward shorter wavelength. Fig.4 shows the result for the objects with $NB497 < 25.0$ and $BV-NB407 > 1.0$, avoiding the objects with relatively low signal to noise ratio. The symmetry index (the ratio) of those with  the characteristic ``strong red and weak blue" profile is strongly concentrated around 0.5 while those of other objects are distributed around the unity (The median is 0.83) with the r.m.s. scatter of $\sim 0.25$. 12 objects with characteristic profile has the value below 0.6 and they indeed show the conspicuous ``strong red and weak blue" profiles.  We applied the Kolmogorov-Smirnov Test to the symmetry distribution of the objects with and without the characteristic profile and the probability that they are drawn from the same population is 7\%. We also changed the magnitude or color range as well as the wavelength region for the index and found that the trend is always seen.

\figurenum{4}
%%%%%%%%%%%%%%%%%%%%%%%%%%%%%%%%%%%%%%%%
\begin{figure}[h]
\includegraphics[width=8cm]{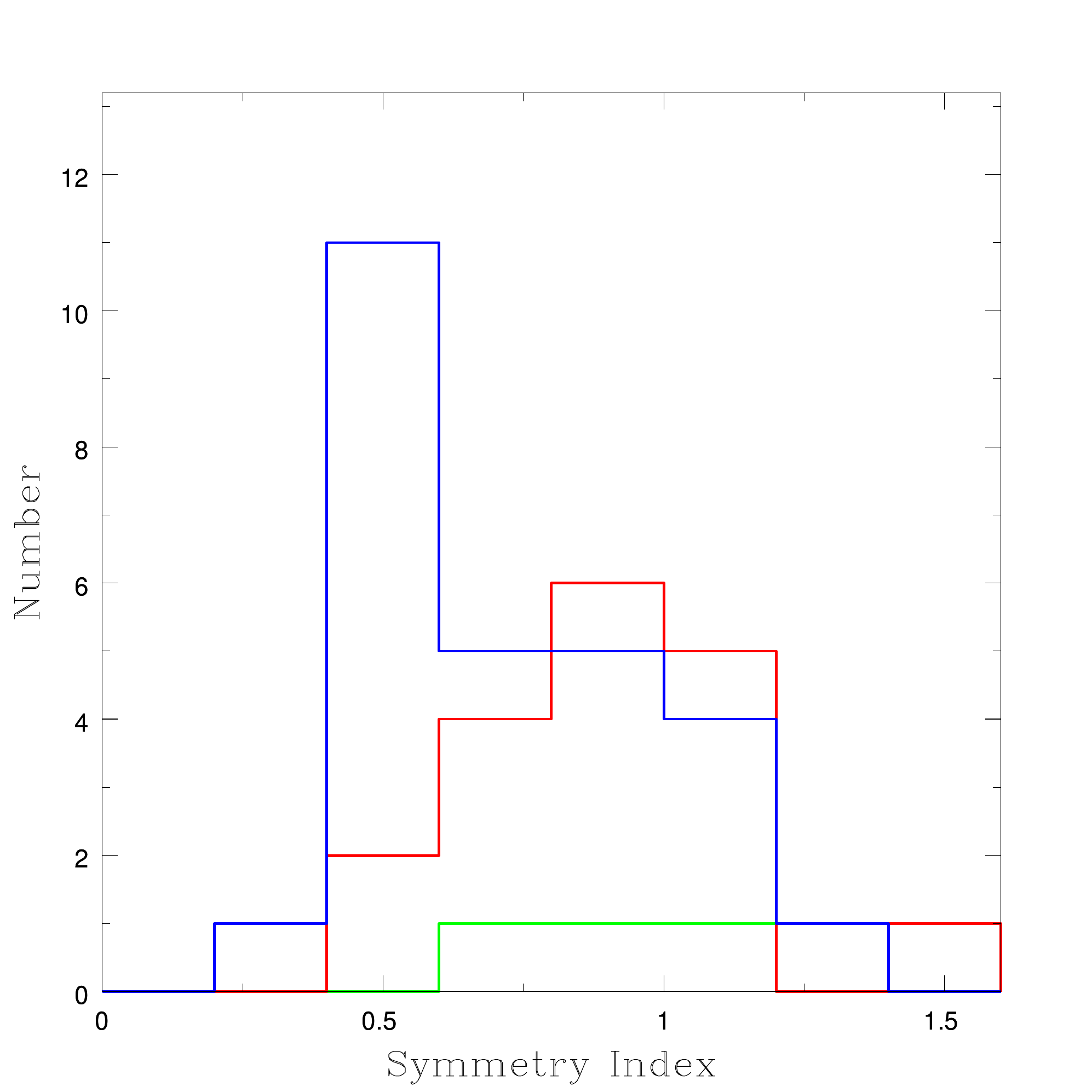}
\caption{The distribution of the symmetry index of the strongest peak in the Ly$\alpha$ profile. The objects with the characteristic profile with strong red and weak blue components are shown by the blue histogram which strongly peaked at 0.5. The objects with strong blue and weak red peaks are shown  by the green line. Other objects are shown by the red line.
}
\label{fig4}
\end{figure}
%%%%%%%%%%%%%%%%%%%%%%%%%%%%%%%%%%%%%%%%

\figurenum{5}
%%%%%%%%%%%%%%%%%%%%%%%%%%%%%%%%%%%%%%%%
\begin{figure}[h]
\includegraphics[width=8cm]{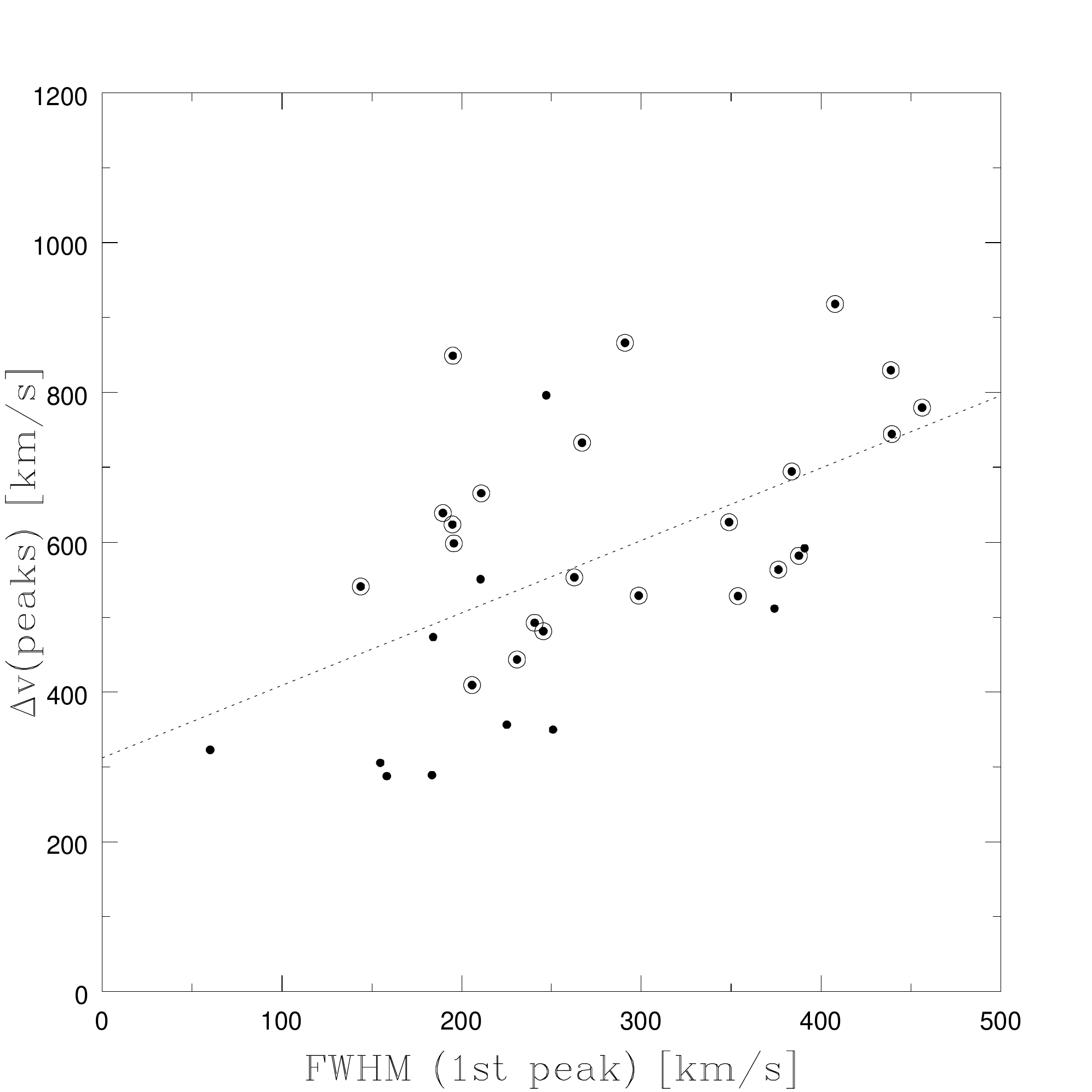}
\caption{The correlation between the observed width of the strongest peak and the separation of the first and the second peaks for the multiple peak objects. Those with relatively good S/N, with $NB497 <25$ and $BV-NB497 >1.0$ are marked by the large open circles. The dotted line is the best fit regression line.}
\label{fig5}
\end{figure}
%%%%%%%%%%%%%%%%%%%%%%%%%%%%%%%%%%%%%%%%

 Next, for those objects with the characteristic profile, we compared the observed width of the stronger red peak and the separation of the peaks obtained in the formal Gaussian fitting. Fig.5 shows the result; the clear correlation between these two quantities is seen. This strongly support that the two peaks are indeed the parts of a single feature. If they are multiple different components, such trend as seen in Fig.5 is not necessary to be observed. Fig.6 shows the distribution of the observed FWHM of the most prominent peak of the each object. Those with the characteristic profile tend to show the smaller values.

\figurenum{6}
%%%%%%%%%%%%%%%%%%%%%%%%%%%%%%%%%%%%%%%%
\begin{figure}[h]
\includegraphics[width=8cm]{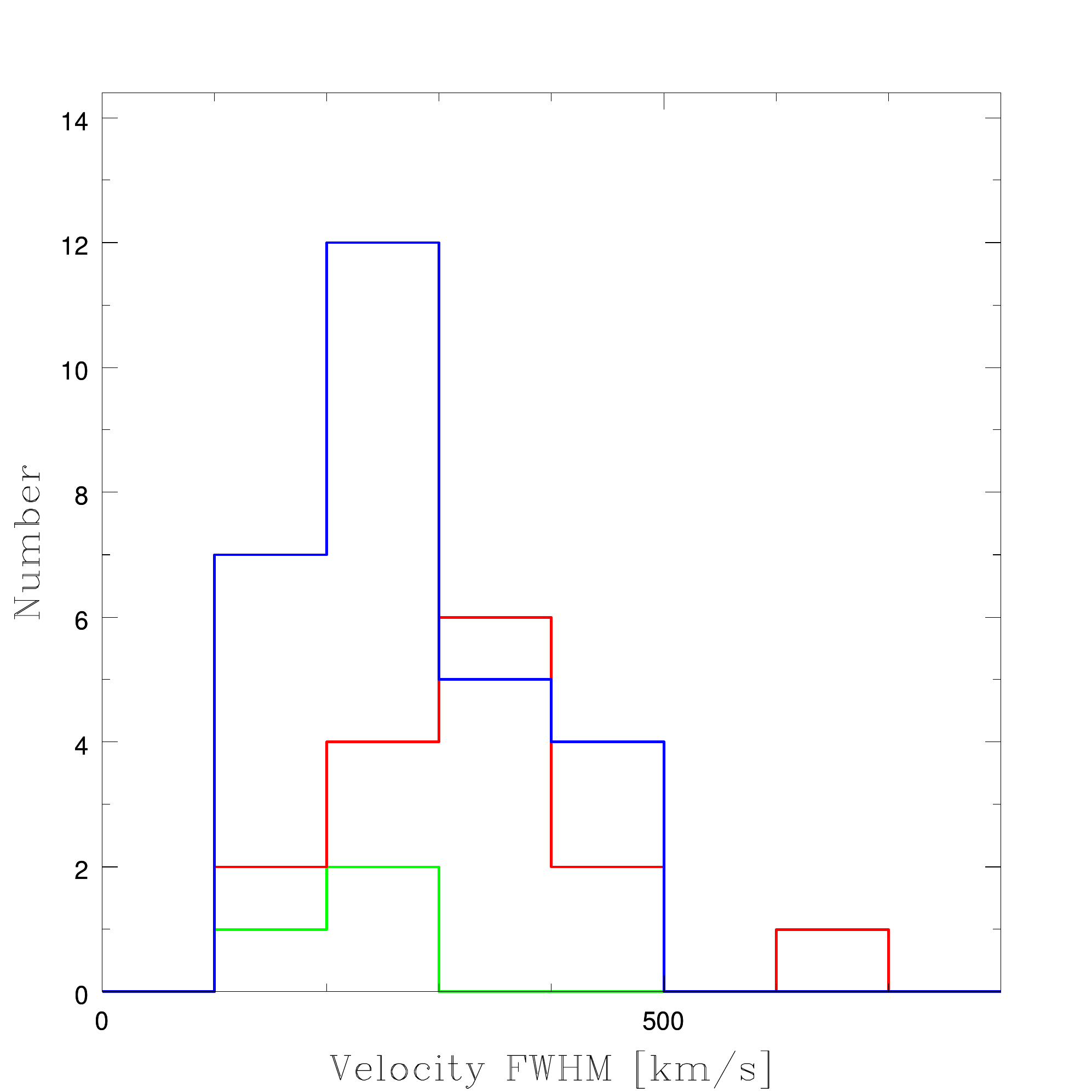}
\caption{The observed velocity FWHM before the instrumental correction for the most prominent peak of each object. Those with relatively good S/N, with $NB497 <25$ and $BV-NB497 >1.0$ are used. Same as in Fig.4, the blue and red line show the objects with multiple characteristic peaks and those with a single peak, respectively.
}
\label{fig6}
\end{figure}
%%%%%%%%%%%%%%%%%%%%%%%%%%%%%%%%%%%%%%%%

\section{Discussions}

\subsection{Characteristic Ly$\alpha$ Line Profile}

 We found that a large fraction, $>$40$\%$ of the observed Ly$\alpha$ emitters in SSA22-Sb1 field have a characteristic double-peaked line profile with a strong red peak and a much weaker blue peak. What does this characteristic Ly$\alpha$ profile means?  The double peaked profiles can be a single source absorbed or scattered by the neutral hydrogen gas or multiple components of the ionized gas with different velocities. The multiple component cases are not favored, however, as the observed lines show asymmetry and there is a correlation between the peak separation and the line width as described in the previous section.
 
 Ly$\alpha$ line profile emission is affected by intergalactic medium (IGM) in the line of sight toward the object (Dijikstra et al. 2007; Laursen et al. 2011). At $z \sim 3$, the mean transmission of the photon at the wavelength just shorter than Ly$\alpha$ due to the absorption by the foreground intergalactic Ly$\alpha$ clouds is $\sim 0.7$ (Madau et al. 1996). Due to the fluctuation by the large-scale structure, the absorption varies by the line of sight toward the sources. Laursen et al. (2011) recently examined how the line profile of Ly$\alpha$ emitter at z=2.5-6.5 is affected by IGM absorption. The case for $z=3.5$ (their Figure 7) indeed predicts the double peaked Ly$\alpha$ line profile with the stronger red component. The red and blue peak-strength difference is more conspicuous for the smaller objects. As the size of the emitter becomes larger, the two peaks show comparable strength on average although the distribution of the IGM optical depth can produce the variation of the profiles. The prediction may be compared with our observed results. For those objects identified to have the characteristic profile, most of them show the large flux difference between the red and blue peaks and there are few objects with comparable strength (Fig.1). The formal single-component Gaussian fitting shows that more than 50\% of the line flux is absorbed for the typical cases (e.g., Fig.2b, Fig.3b). It seems difficult to explain the current results only by IGM absorption in this sense. Note that, however, the emitters observed here are located in the dense environment and the IGM absorption is much larger than average. We revisit this possibility in Sec.4.4.

 We then consider the cases that the profiles show the intrinsic properties of the Ly$\alpha$ emitters, namely the effects by the gas associated with them. Scattering in the infalling partially ionized emission-line gas (Dijikstra et al. 2006) as well as scattering in the expanding shell-like out-flowing gas (Verhamme et al. 2006; 2008) can modify the line profile. For the case of spherically symmetric infalling gas model, Dijikstra et al. (2006) argued that the entire line profile shows strong blue symmetry and the blue peaks are observed stronger than the red peaks. As the red peaks are much stronger than the blue peak for the most of the cases studied here, the other interpretation, namely the expanding shell model is more favored. For the spherically symmetrical expanding shell model (Verhamme et al. 2006) surrounding the Ly$\alpha$ source, the Ly$\alpha$ line is shifted red ward from the systemic velocity and also show strong red asymmetry. Indeed, some model profiles shown in Verhamme et al. (2006) (their Fig.15) as well as in Verhamme et al. (2008) seems quite similar with the characteristic ``strong red and weak blue" profiles discussed here. Recently, McLinden et al. (2011) reported that a strong [OIII] 5007\AA\ emission line is detected for a Ly$\alpha$ emitter at $z=3.1$ which show the similar double peaked Ly$\alpha$ profile. Interestingly, the [OIII] line locates at the middle of the two Ly$\alpha$ peaks, which implies that the systemic velocity of the galaxy is indeed at the middle of the two Ly$\alpha$ peaks. Thus the expanding shell by Verhamme et al. (2008) is favored to understand the observed line profiles. In this case, the star-formation in these Ly$\alpha$ emitters must be dominated by the burst-like events since the continuous gas infall is needed to maintain continuous star-formation activities. The population of the Ly$\alpha$ emitters are generally considered to be less massive objects in stellar mass as the many of them are not detected even in the deepest ground-based near-infrared images. The early active starburst events may result in the galactic wind activity producing the expanding shells. It is important to note that the Ly$\alpha$ emission from $\sim 44$$\%$, and possibly a larger fraction of the observed sample, show such evidence. 
 
 To prove the expanding shell or gas outflow models, another important observational constraint is the surface-brightness distribution of the Ly$\alpha$ emission. Rauch et al (2008), having obtained 2-dimensional spectra of Lyman alpha emitters and found the surface-brightness to be mostly strongly peaked in the spatial direction, even though the emission could be traced out to several arcseconds. Barnes \& Haehnelt (2010) argued that this rapid drop of surface brightness with radius is inconsistent with the much flatter radial profile predicted by a simple expanding shell model. In fact, there is a weak trend, though not very significant, that the objects with the characteristic profile are rather small (see Sec.4.3 and Fig.8 and Fig.9 below). This is against the interpretation by the simple expanding-shell models. For much extended objects such as Ly$\alpha$ Blobs, on the other hand, Mori and Umemura (2006) successfully reproduce the observed Ly$\alpha$ morphology by the models of expanding gas heated by supernovae. Future improvements both in observations and models (e.g., Barnes et al. 2011) will allow more detailed study on the surface-brightness distributions.

\subsection{Lyman $\alpha$ Blobs}

 It is also interesting to discuss the relation between characteristic line profile and the Ly$\alpha$ Blobs. Matsuda et al. (2006) found that the galaxies with larger size tend to have more structures in the line profiles and 'total line width' larger than FWHM $\sim 500$ km s$^{-1}$ if fitted by a single Gaussian profile (as in Fig.2b or 3b in this paper). 

 For the velocity width, a similar trend is confirmed with the current sample. In Table 3, we listed the velocity FWHM obtained by the single Gaussian fitting, namely the same procedure as in Matsuda et al. (2006), for the 12 objects associated with the Ly$\alpha$ Blobs in Matsuda et al. (2004). Of the 12 objects, 9 have FWHM $> 450$ km s$^{-1}$. While the rest three objects have the smaller velocity width, LAB32 and LAB33 are among the smallest in the sample of the 35 Ly$\alpha$ Blobs in Matsuda et al. (2004). LAB7 has three, or possibly four distinguished Ly$\alpha$ peaks (Matsuda et al. 2004) and the slit position in this study just covers a single south-most peak and the value measured in this observation may not represent the whole nature of the system. 
 
 Of the 12 objects associated with the Ly$\alpha$ Blobs, on the other hand, only 5 are classified as the ``strong red and weak blue" profile objects. At a glance, this appears somewhat strange, since the line profiles of the Ly$\alpha$ Blobs shown in Matsuda et al. (2006) have more structures with possible absorption than the other smaller emitters. If we closely see Fig.1, however, there are hints of the multiple components which are not significantly detected in the current data. Although it is difficult to conclude, as the exposure time (7 hours in Matsuda et al. 2006), spectral resolution ($R \sim 2500$), instrument ($Keck$ DEIMOS) are different from those in this paper, the deeper data may probe significant structures of these lines. Other possible interpretation is as follows. If Ly$\alpha$ Blobs are more massive objects, they may have more thick absorbing material and the weak blue line is too much absorbed to be detected. This is likely to be the case for those with the single but asymmetrical component, such as LAE J221808.3+001022 (LAB15), LAE J221812.5+001433 (LAB33), and LAE J221817.3+001209 (LAB21).

 Matsuda et al. (2006) argued that the large velocity width implies that Ly$\alpha$ Blobs are more massive objects than other 'normal' Ly$\alpha $ emitters. While the large size of the Ly$\alpha$ emission of the blobs can be explained by either the expanding gas ionized and excited by the supernova feedback (Mori and Umemura 2006) or by the infalling cooling gas (Haiman et al. 2000; Fardal et al. 2001).  On the other hand, recent models of the cold flows can reproduce the size and luminosity distribution of the Ly$\alpha$ blobs (Dekel et al. 2009; Goerdt et al. 2009). If the Ly$\alpha$ blobs are the scaled-up version of the normal emitters whose line profiles are dominated by the outflowing gas, the gas-outflow interpretation may be favored.

\figurenum{7}
%%%%%%%%%%%%%%%%%%%%%%%%%%%%%%%%%%%%%%%%
\begin{figure}[h]
\includegraphics[width=8cm]{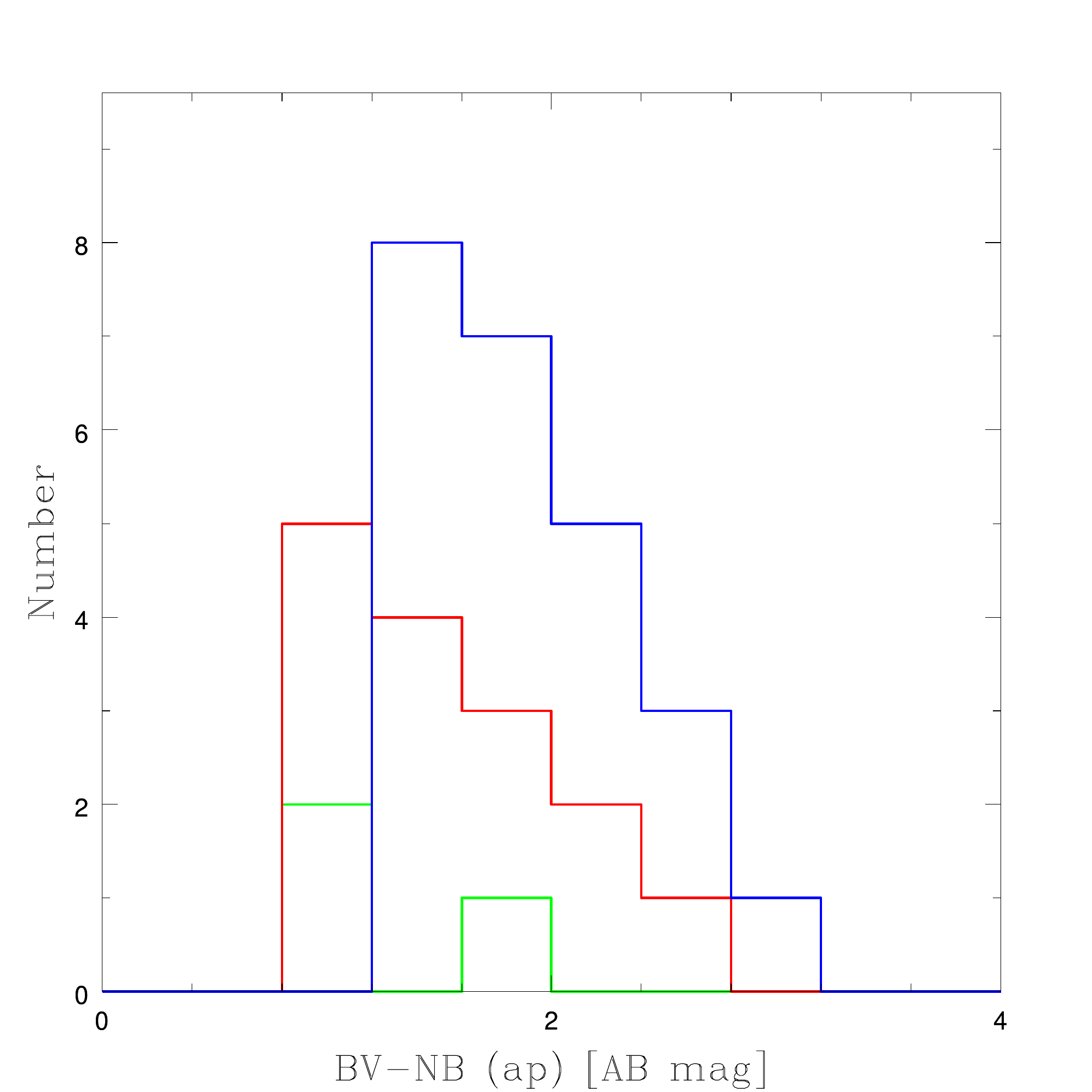}
\includegraphics[width=8cm]{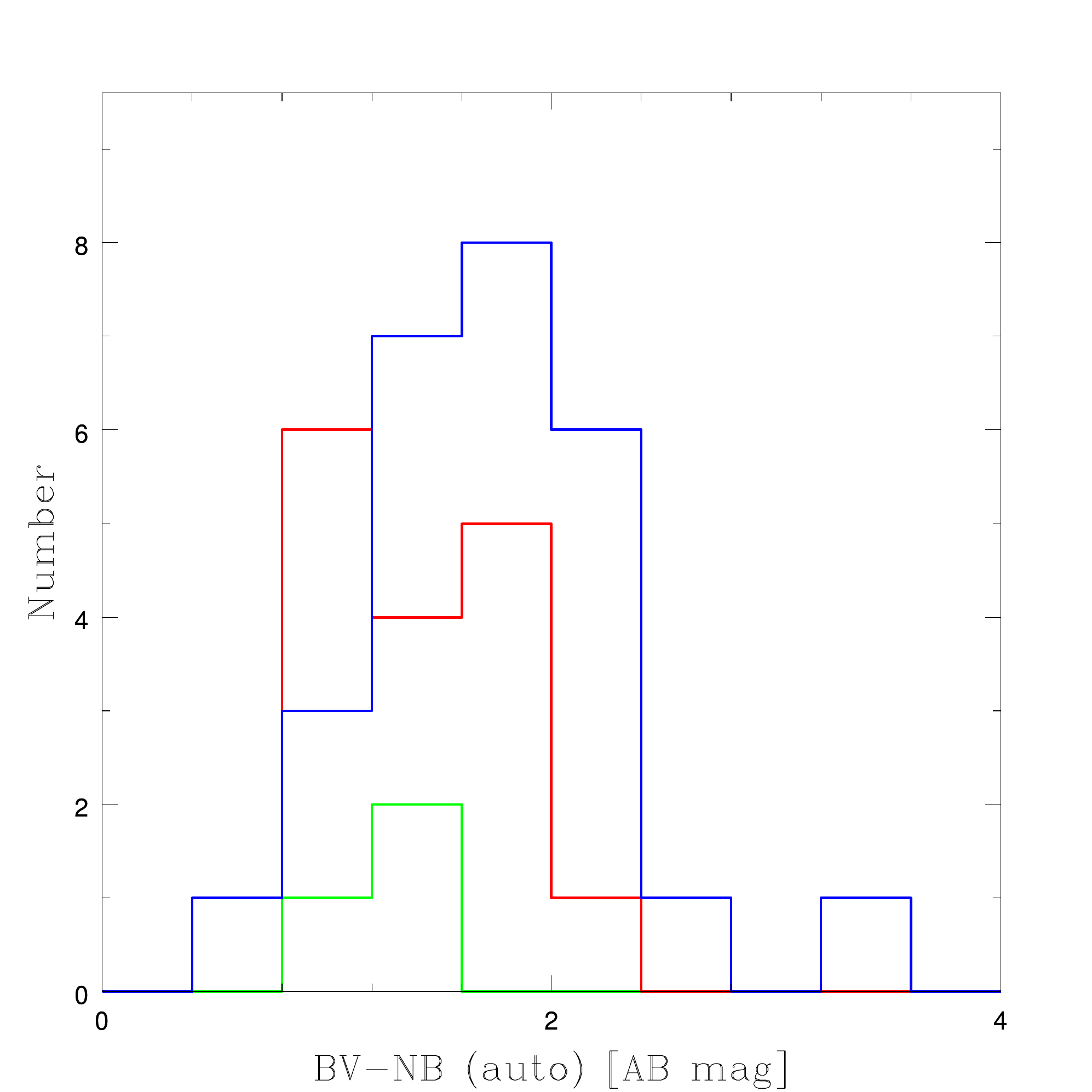}
\caption{The distribution of BV-NB497 color, namely the observed excess in the narrow band. The panel (a) shows the values measured in 2-arcsec diameter aperture while the panel (b) shows those for the MAG AUTO semi-total magnitude. The lines have the same meaning as in Fig.6.
}
\label{fig7}
\end{figure}
%%%%%%%%%%%%%%%%%%%%%%%%%%%%%%%%%%%%%%%%

\subsection{Correlation with other properties}

 We also investigated whether the presence of the characteristic profile is related with any other properties of the Ly$\alpha$ emitters. First, we see the distribution of the observed excess in $BV-NB497$ color, or equivalent width, in Fig.7a and 7b. Fig.7a shows the result for the color measured in the fixed 2$''$-diameter aperture and the Fig.7b shows those for the semi-total flux measured in the Kron aperture (i.e., SExtracter MAG\_AUTO) for those LAEs with $NB497<25.0$ mag.  The emitters with the characteristic profile seem to dominate at $BV-NB497 > 1.8$, where 13/16 objects show ``strong red and weak blue" peaks, although the $BV-NB497$ colors excess of the objects with the characteristic profile is not always large. The trend is also notable in Fig.7b. The probabilities for Kolmogorov-Smirnov Test that they follow the same distribution are 17$\%$ and $9\%$, respectively. The large excess in the narrow band, or Ly$\alpha$ equivalent width can be related to the gas outflow such as additional Ly$\alpha$ emission due to the collisional excitation or ionization by the shock. If the source of the Ly$\alpha$ emission is the photoionization by massive stars, the large equivalent width suggests the presence of a very young or metal-poor stellar population, which is also consistent with the gas outflow by the supernovae or strong stellar wind. 
 
   We also studied the distributions of the size, observed FWHM and the isophotal area above the detection threshold and the results are shown in Fig.8 and Fig.9. To omit the objects with low S/N ratio, we again limited the sample to those with $NB497 < 25.0$ and $BV-NB497 > 1.0$. While the distributions for the objects with and without the characteristic profile are not quite distinguishable, there is some concentration of the ``strong red and blue weak" object for smaller sizes (Kolmogorov-Smirnov Test probability is $15\%$). This is disadvantageous for the simple expanding shell model which predicts a rather flat surface brightness distribution. If the starburst is so young, however, the area of the outflowing gas can be still very limited within small radius from the galactic center to show nearly unresolved morphology.

\figurenum{8}
%%%%%%%%%%%%%%%%%%%%%%%%%%%%%%%%%%%%%%%%
\begin{figure}[h]
\includegraphics[width=8cm]{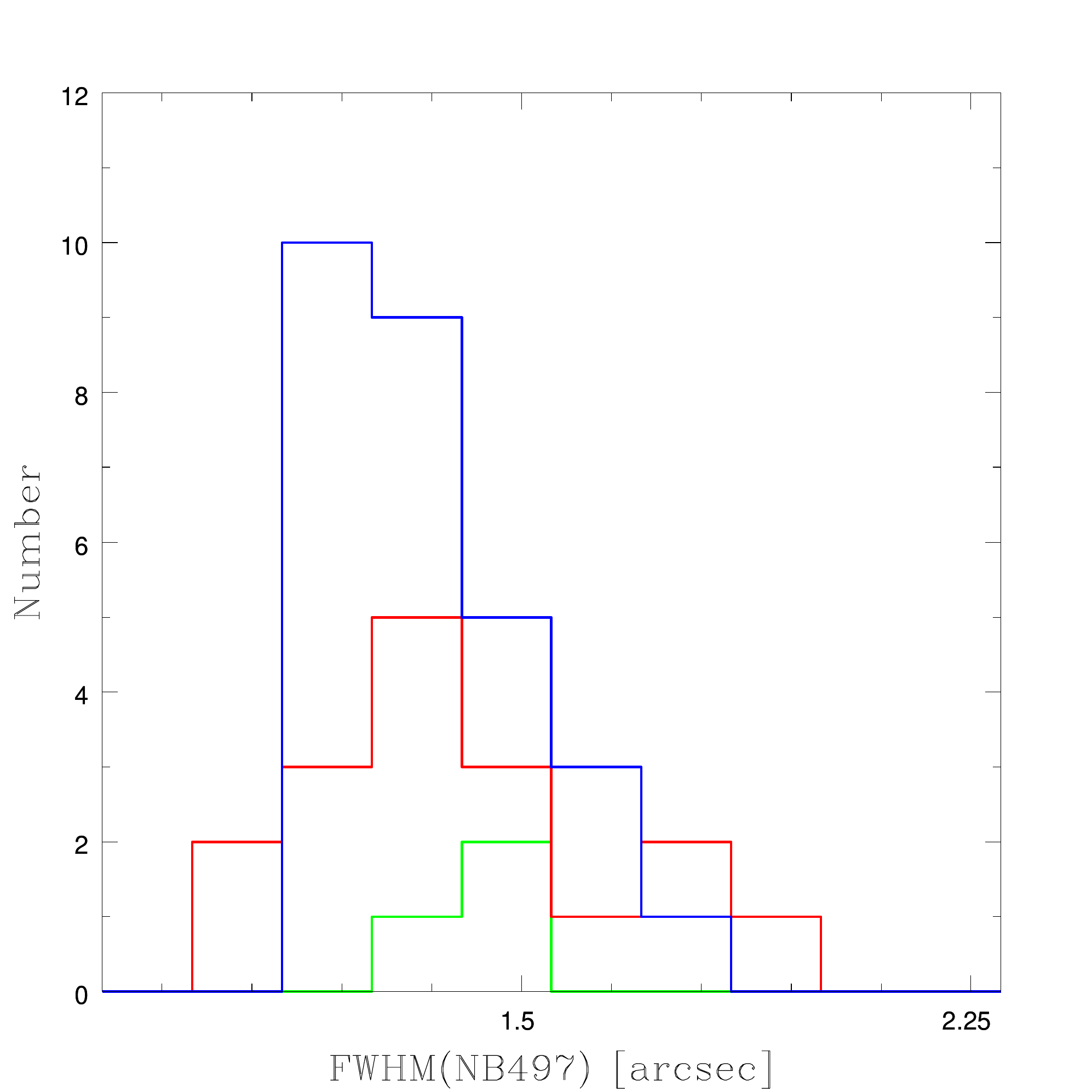}
\caption{The distribution of the size of the L$\alpha$ emitters in FWHM in their surface brightness profile. The lines have the same meaning as in Fig.6.
}
\label{fig8}
\end{figure}
%%%%%%%%%%%%%%%%%%%%%%%%%%%%%%%%%%%%%%%%

\figurenum{9}
%%%%%%%%%%%%%%%%%%%%%%%%%%%%%%%%%%%%%%%%
\begin{figure}[h]
\includegraphics[width=8cm]{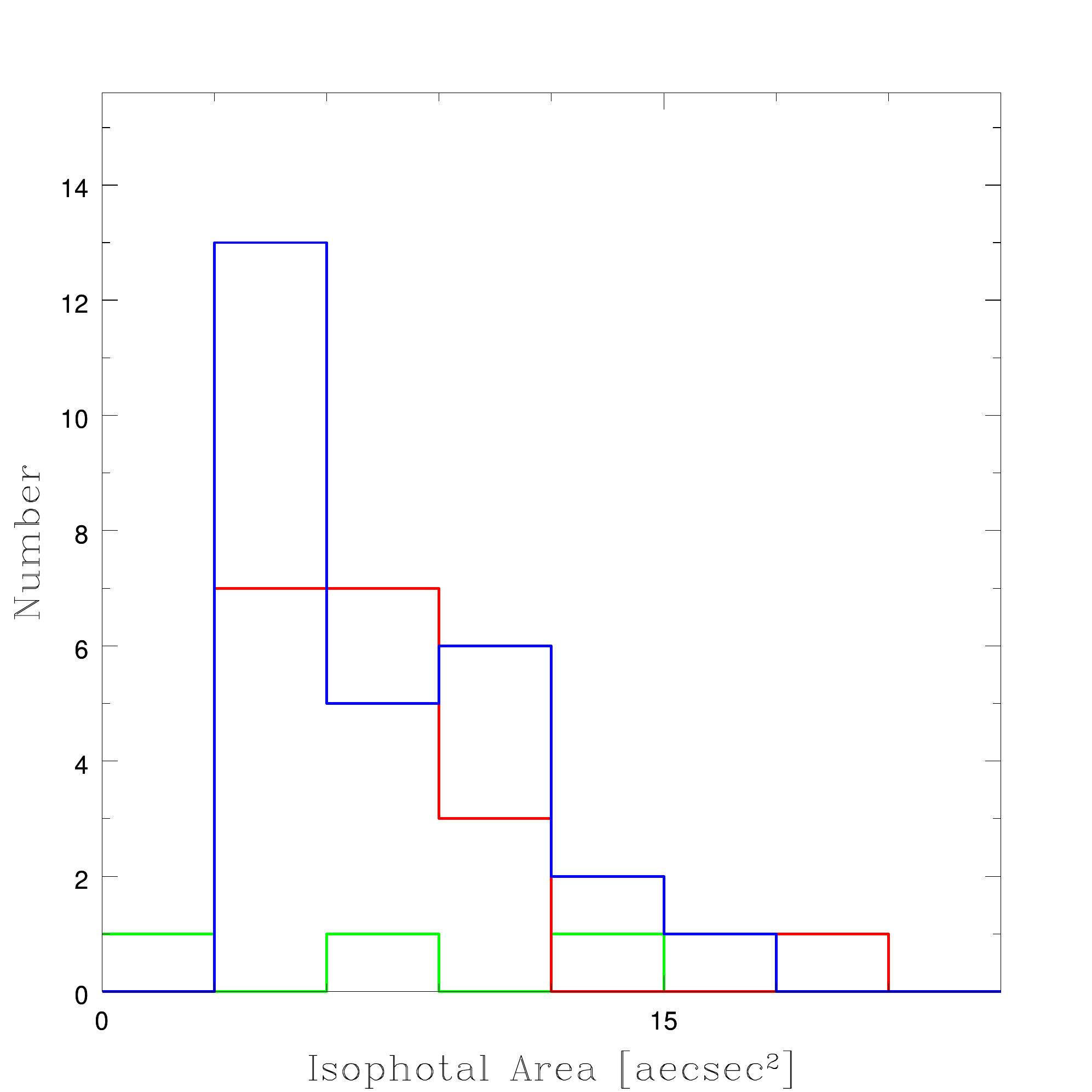}
\caption{The distribution of the narrow-band isophotal area above the detection threshold. The lines have the same meaning as in Fig.6.
}
\label{fig9}
\end{figure}
%%%%%%%%%%%%%%%%%%%%%%%%%%%%%%%%%%%%%%%%

\subsection{Intergalactic Absorption in Protocluster?}

 Finally, we revisit the possibility that the characteristic profile is caused by the intergalactic absorption systems. As the field is known to have a high density of star-forming galaxies (Steidel et al. 1998; 2000, Hayashino et al. 2004), the neutral gas density may also be enhanced along the filamentary large-scale structure. The characteristic profile can be caused by the absorption by the common gas in such structure in front of the observed emitters. In such a case, absorption redshift is expected to be spatially correlated. Fig.10 shows the sky distribution of the Ly$\alpha$ emitters with the characteristic profile. The redshift of the ``absorption" that was formally obtained by the multiple component Gaussian fitting allowing the absorption features are color coded, from the redshift 3.047 to 3.129 with the interval of 0.008 (600 km s$^{-1}$). No large-scale correlation of the absorption redshift is seen.

\figurenum{10}
%%%%%%%%%%%%%%%%%%%%%%%%%%%%%%%%%%%%%%%%
\begin{figure}[h]
\includegraphics[width=8cm]{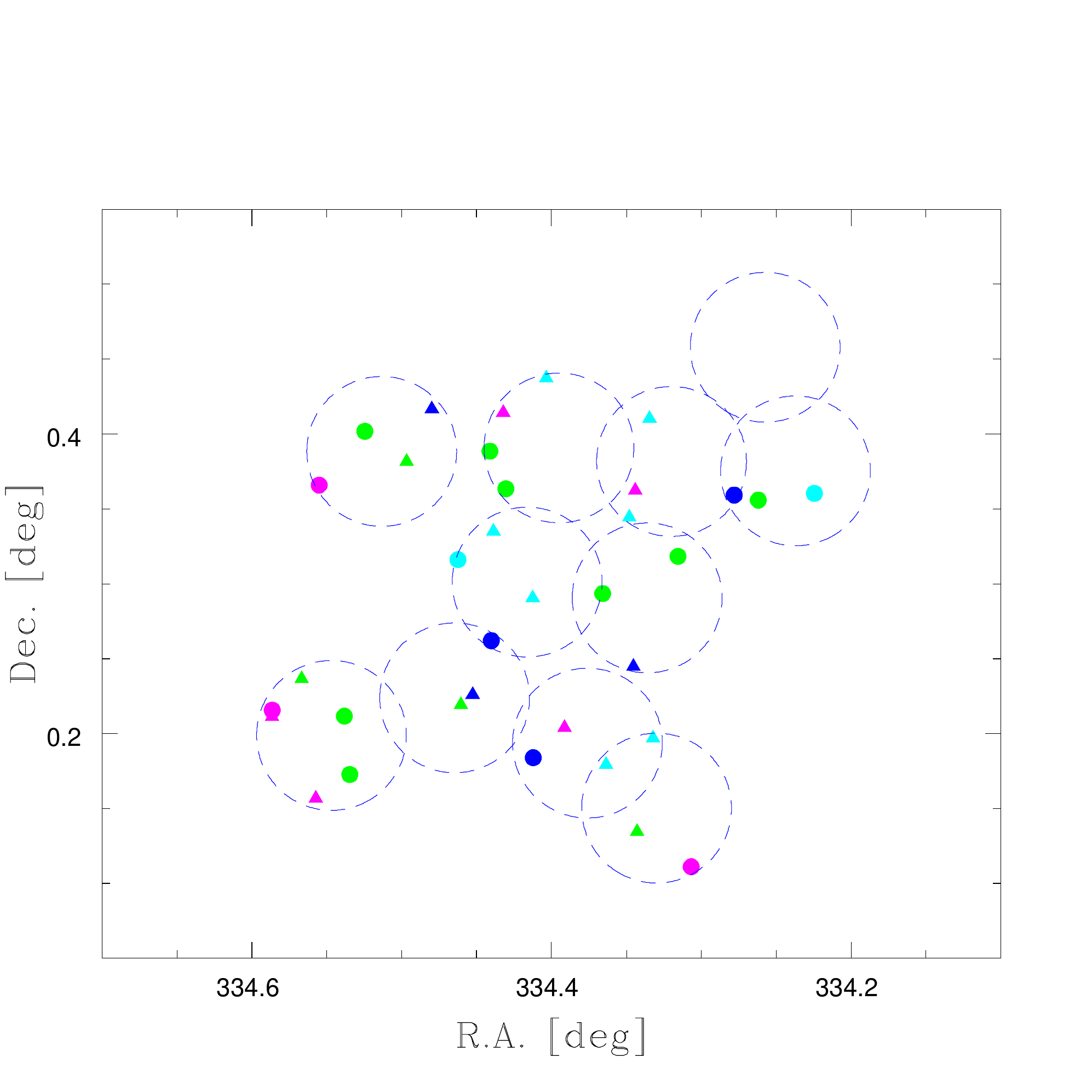}
\caption{The sky distribution of the objects with the characteristic profiles. The dashed large circle shows the FOCAS slit mask areas. The objects with the different symbols and colors have the different redshift for the absorption in the formal Gaussian fitting. Redshift increases from 3.047 to 3.129 with the interval of 0.008, in order of blue (circle, triangle), cyan, green, magenta, and red colors.}
\label{fig10}
\end{figure}
%%%%%%%%%%%%%%%%%%%%%%%%%%%%%%%%%%%%%%%%

\vspace*{5mm}

 We thank the staff of the Subaru Telescope for their assistance. This research is supported in part by the Grant-in-Aid 20450224 for Scientific Research of the Ministry of Education, Science, Culture, and Sports in Japan. Data analysis was in part carried out on common use data analysis computer system at the Astronomy Data Center, ADC, of the National Astronomical Observatory of Japan.

%%%%%%%%%%%%%%%%%%%%%%%%%%%%%%%%%%%
%%%%%%%%%%%%%%%%%%%%%%%%%%%%%%%%%%%

%%%%%%%%%%%%%%%%%%%%%%%%%%%%%%%%%%%%%%%
%%%%%%%%%%%%%%%%%%%%%%%%%%%%%%%%%%%%%%%

\clearpage

\tablenum{1}

%%%%%%%%% Table %%%%%%%%
\begin{table*}
\begin{center}
\caption{Summary of the Observed Objects\label{tab1}}
\begin{tabular}{ccccccccc}
\tableline\tableline
Name &    z$^a$ &  $\Delta V_{\rm FWHM corr}^b$ & $\lambda_1^c$ &  $\lambda_2^d$ &  $\lambda_3^e$  &   NB$_{\rm auto}$ &   NB$_{\rm ap}^f$ &   BV$_{\rm ap}^f$  \\ 
     &      & km s$^{-1}$ &  \AA\       & \AA\   &  \AA\     &   &  &  \\

\tableline
LAE J221650.2+002750 &   3.099 &     171.1 &   4984.9 &      0.0 &   4984.9 &    25.24 &    25.68 &    26.88  \\ 
LAE J221650.9+002327 &   3.087 &     191.1 &   4970.8 &      0.0 &   4970.8 &    25.00 &    25.14 &    26.31  \\ 
LAE J221652.7+002423 &   3.067 &     217.3 &   4944.7 &      0.0 &   4944.7 &    24.73 &    24.89 &    25.89  \\ 
LAE J221653.9+002137 &   3.081 &     402.5 &   4964.6 &   4950.9 &   4966.6 &    24.57 &    24.67 &    26.27  \\ 
LAE J221658.4+002430 &   3.102 &     392.0 &   4987.1 &      0.0 &   4987.1 &    24.07 &    24.61 &    25.90  \\ 
LAE J221659.3+002304 &   3.069 &     306.0 &   4947.2 &      0.0 &   4947.2 &    23.95 &    24.92 &    25.69  \\ 
LAE J221659.3+002501 &   3.088 &     141.5 &   4970.7 &   4960.9:&   4969.3 &    24.30 &    24.97 &    26.08  \\ 
LAE J221700.3+002404 &   3.098 &      96.7 &   4981.8 &      0.0 &   4981.8 &    24.61 &    24.78 &    26.35  \\ 
LAE J221700.7+002006 &   3.075 &      42.5 &   4955.3 &      0.0 &   4956.4 &    25.17 &    25.22 &    26.66  \\ 
LAE J221701.6+002135 &   3.070 &     265.8 &   4947.6 &   4932.9 &   4942.2 &    24.59 &    24.96 &    25.63  \\ 
LAE J221701.8+002003 &   3.077 &     161.9 &   4956.1 &   4947.0 &   4953.1 &    25.21 &    25.25 &    28.03  \\ 
LAE J221702.7+002433 &   3.067 &     385.4 &   4944.7 &      0.0 &   4944.7 &    23.58 &    23.88 &    24.91  \\ 
LAE J221702.8+002121 &   3.091 &       0.0 &   4973.5 &   4968.7 &   4971.7 &    25.27 &    25.34 &    27.70  \\ 
LAE J221704.9+002226 &   3.108 &     279.8 &   4995.5 &      0.0 &   4995.5 &    24.80 &    25.26 &    26.63  \\ 
LAE J221706.7+002134 &   3.067 &     330.4 &   4944.2 &   4935.8 &   4941.9 &    23.60 &    24.11 &    24.97  \\ 
LAE J221709.6+001801 &   3.105 &     924.1 &   4994.2 &      0.0 &   4987.8 &    23.07 &    23.08 &    24.75  \\ 
LAE J221713.3+002034 &   3.101 &     278.0 &   4987.2 &      0.0 &   4987.2 &    24.73 &    25.06 &    27.55  \\ 
LAE J221713.6+000640 &   3.113 &     302.7 &   5002.0 &   4991.5 &   5000.9 &    23.23 &    23.31 &    24.82  \\ 
LAE J221713.7+001656 &   3.057 &     249.4 &   4934.2 &      0.0 &   4934.2 &    24.75 &    25.13 &    26.45  \\ 
LAE J221715.7+001906 &   3.101 &     233.2 &   4986.7 &   4972.3 &   4986.0 &    24.55 &    24.60 &    27.03  \\ 
LAE J221715.8+002431 &   3.106 &     276.5 &   4992.4 &      0.0 &   4992.4 &    24.47 &    24.68 &    26.40  \\ 
LAE J221716.7+002309 &   3.066 &     388.7 &   4945.7 &      0.0 &   4945.7 &    24.53 &    25.36 &    26.61  \\ 
LAE J221717.3+000728 &   3.090 &     149.3 &   4972.9 &      0.0 &   4972.9 &    24.47 &    25.18 &    26.53  \\ 
LAE J221717.5+002010 &   3.075 &     563.1 &   4954.4 &      0.0 &   4954.7 &    24.65 &    25.36 &    28.47  \\ 
LAE J221718.8+001518 &   3.067 &      91.1 &   4944.8 &      0.0 &   4943.4 &    24.44 &    24.90 &    26.98  \\ 
LAE J221719.0+001200 &   3.090 &     265.7 &   4972.1 &   4965.6:&   4970.3 &    25.03 &    25.13 &    26.64  \\ 
LAE J221719.3+001450 &   3.065 &     288.8 &   4942.8 &   4935.2:&   4943.8 &    24.40 &     0.00 &     0.00  \\ 
LAE J221719.7+001149 &   3.067 &     150.1 &   4945.4 &   4938.1 &   4943.4 &    24.83 &     0.00 &     0.00  \\ 
LAE J221720.1+002226 &   3.104 &     226.7 &   4991.1 &      0.0 &   4991.1 &    24.87 &    25.40 &    26.73  \\ 
LAE J221720.2+002019 &   3.108 &    1997.2 &   5005.8 &      0.0 &   4995.0 &    22.44 &    22.40 &    23.43  \\ 
LAE J221720.3+002438 &   3.070 &     201.4 &   4948.0 &   4935.9 &   4943.9 &    23.59 &    23.87 &    25.18  \\ 
\tableline
\end{tabular}
\end{center}
\end{table*}

\tablenum{1}

%%%%%%%%% Table %%%%%%%%
\begin{table}
\begin{center}
\caption{Continued. \label{tab1}}
\begin{tabular}{ccccccccc}
\tableline\tableline
Name &    z$^a$ &  $\Delta V_{\rm FWHM corr}^b$ & $\lambda_1^c$ &  $\lambda_2^d$ &  $\lambda_3^e$  &   NB$_{\rm auto}$ &   NB$_{\rm ap}^f$ &   BV$_{\rm ap}^f$  \\ 
     &      & km s$^{-1}$ &  \AA\       & \AA\   &  \AA\     &   &  &  \\
\tableline\tableline
LAE J221721.7+001223 &   3.068 &     355.1 &   4947.0 &      0.0 &   4947.0 &    25.06 &    25.32 &    26.90  \\ 
LAE J221722.3+000804 &   3.085 &      88.4 &   4967.3 &   4957.4 &   4962.0 &    23.81 &    24.07 &    25.55  \\ 
LAE J221722.6+002145 &   3.104 &      88.1 &   4988.4 &   4978.1 &   4989.0 &    25.00 &    25.05 &    27.06  \\ 
LAE J221722.9+001441 &   3.054 &       0.0 &   4929.2 &   4924.1 &   4928.7 &    24.99 &    25.08 &    26.90  \\ 
LAE J221723.5+002040 &   3.071 &     368.3 &   4950.8 &   4935.7 &   4950.9 &    24.76 &    24.88 &    26.51  \\ 
LAE J221723.8+002155 &   3.102 &       0.0 &   4987.2 &      0.0 &   4987.2 &    24.71 &    25.49 &    27.09  \\ 
LAE J221724.6+001557 &   3.079 &     172.0 &   4960.1 &   4953.5:&   4958.8 &    25.34 &    25.42 &    28.47  \\ 
LAE J221724.7+002227 &   3.091 &     349.8 &   4974.4 &      0.0 &   4975.2 &    25.12 &    25.37 &    26.75  \\ 
LAE J221724.8+001717 &   3.096 &     358.3 &   4981.2 &      0.0 &   4980.8 &    23.62 &     0.00 &     0.00  \\ 
LAE J221727.2+001622 &   3.096 &     339.5 &   4978.1 &      0.0 &   4978.1 &    24.67 &    24.81 &    26.33  \\ 
LAE J221727.3+001046 &   3.069 &     307.2 &   4946.9 &   4938.1 &   4945.3 &    24.64 &    24.77 &    27.02  \\ 
LAE J221727.8+001737 &   3.092 &     172.9 &   4975.6 &   4967.6 &   4973.2 &    24.31 &    24.40 &    26.97  \\ 
LAE J221728.3+001212 &   3.067 &     162.3 &   4945.3 &      0.0 &   4947.5 &    24.31 &    24.51 &    26.27  \\ 
LAE J221733.9+001215 &   3.106 &       0.0 &   4991.5 &   4982.5 &   4990.7 &    24.67 &    25.04 &    26.93  \\ 
LAE J221735.9+001559 &   3.094 &     673.8 &   4978.0 &      0.0 &   4978.0 &    24.08 &    24.68 &    26.92  \\ 
LAE J221736.4+001251 &   3.058 &     150.5 &   4934.1 &      0.0 &   4934.1 &    25.20 &    25.17 &    27.31  \\ 
LAE J221736.8+002614 &   3.069 &     179.5 &   4948.0 &   4942.2 &   4947.3 &    24.98 &    25.06 &    27.40  \\ 
LAE J221738.5+002215 &   3.096 &     144.4 &   4980.3 &      0.0 &   4980.3 &    24.85 &    24.96 &    26.74  \\ 
LAE J221738.9+001102 &   3.064 &     116.9 &   4940.9 &   4929.9 &   4937.3 &    23.56 &    24.06 &    25.36  \\ 
LAE J221739.0+001726 &   3.072 &     165.0 &   4951.5 &   4943.3 &   4949.0 &    24.66 &     0.00 &     0.00  \\ 
LAE J221739.2+002242 &   3.098 &     189.3 &   4981.4 &      0.0 &   4983.8 &    25.25 &    25.52 &    27.64  \\ 
LAE J221739.3+001610 &   3.093 &      66.0 &   4976.5 &      0.0 &   4976.4 &    24.84 &    25.40 &    27.51  \\ 
LAE J221739.9+002142 &   3.089 &     361.1 &   4972.8 &      0.0 &   4972.8 &    25.24 &    25.21 &    27.66  \\ 
LAE J221740.3+001129 &   3.065 &      71.9 &   4941.4 &      0.0 &   4943.2 &    25.01 &    24.58 &    25.43  \\ 
LAE J221740.9+001125 &   3.089 &     288.4 &   4968.8 &      0.0 &   4968.8 &    23.49 &    24.53 &    25.31  \\ 
LAE J221741.4+002227 &   3.096 &     208.0 &   4980.9 &      0.0 &   4980.9 &    24.31 &    24.55 &    25.35  \\ 
LAE J221743.3+002149 &   3.097 &      88.2 &   4981.9 &   4967.8 &   4975.4 &    24.68 &     0.00 &     0.00  \\ 
LAE J221743.4+001348 &   3.098 &     142.8 &   4982.1 &   4975.9 &   4980.9 &    23.95 &    24.01 &    25.97  \\ 
LAE J221743.7+002451 &   3.101 &     350.0 &   4987.4 &   4977.5 &   4984.3 &    25.04 &    25.18 &    26.91  \\ 
LAE J221745.3+002006 &   3.074 &     195.8 &   4953.0 &   4943.9 &   4950.3 &    24.00 &    24.13 &    26.53  \\ 
\tableline
\end{tabular}
\end{center}
\end{table}

\tablenum{1}

%%%%%%%%% Table %%%%%%%%
\begin{table}
\begin{center}
\caption{Continued. \label{tab1}}
\begin{tabular}{ccccccccc}
\tableline\tableline
Name &    z$^a$ &  $\Delta V_{\rm FWHM corr}^b$ & $\lambda_1^c$ &  $\lambda_2^d$ &  $\lambda_3^e$  &   NB$_{\rm auto}$ &   NB$_{\rm ap}^f$ &   BV$_{\rm ap}^f$  \\ 
     &      & km s$^{-1}$ &  \AA\       & \AA\   &  \AA\     &   &  &  \\
\tableline
LAE J221745.6+001544 &   3.065 &     241.6 &   4942.3 &   4933.6 &   4939.4 &    24.58 &    24.59 &    26.25  \\ 
LAE J221745.9+002319 &   3.098 &     142.8 &   4981.9 &   4976.0 &   4979.9 &    24.81 &     0.00 &     0.00  \\ 
LAE J221748.6+001334 &   3.056 &      93.5 &   4932.7 &   4924.5 &   4930.5 &    23.90 &     0.00 &     0.00  \\ 
LAE J221750.5+001310 &   3.089 &      57.0 &   4970.6 &   4965.9 &   4967.5 &    25.52 &    25.35 &    27.81  \\ 
LAE J221751.0+001858 &   3.081 &     346.1 &   4963.7 &   4954.0 &   4961.6 &    24.17 &    24.42 &    27.00  \\ 
LAE J221753.2+001238 &   3.095 &     112.4 &   4977.9 &   4969.3:&   4975.1 &    24.60 &     0.00 &     0.00  \\ 
LAE J221754.3+001224 &   3.093 &     184.0 &   4976.4 &      0.0 &   4973.9 &    24.49 &    24.50 &    26.00  \\ 
LAE J221755.1+002460 &   3.058 &     402.7 &   4935.7 &   4923.4 &   4932.3 &    23.73 &    24.03 &    25.33  \\ 
LAE J221757.6+001204 &   3.093 &     282.1 &   4976.6 &   4966.1:&   4973.9 &    24.38 &    24.80 &    26.15  \\ 
LAE J221759.2+002254 &   3.087 &     333.3 &   4968.2 &   4958.8 &   4965.1 &    23.00 &    23.34 &    25.43  \\ 
LAE J221759.3+001148 &   3.055 &      19.4 &   4928.9 &      0.0 &   4929.0 &    25.41 &    25.01 &    25.86  \\ 
LAE J221802.2+002556 &   3.081 &       0.0 &   4961.5 &      0.0 &   4960.3 &    23.03 &    23.88 &    25.97  \\ 
LAE J221803.6+002247 &   3.079 &     133.8 &   4958.9 &      0.0 &   4958.9 &    25.44 &    25.42 &    27.50  \\ 
LAE J221805.9+002407 &   3.095 &       0.0 &   4979.0 &   4973.7 &   4977.0 &    25.24 &    25.38 &    27.25  \\ 
LAE J221807.9+002317 &   3.088 &     131.8 &   4970.1 &   4955.8 &   4964.9 &    23.97 &    24.58 &    26.12  \\ 
LAE J221808.0+001151 &   3.096 &     159.7 &   4980.1 &      0.0 &   4980.1 &    24.82 &    25.19 &    26.33  \\ 
LAE J221808.3+001022 &   3.097 &     421.7 &   4982.6 &   4969.6 &   4980.0 &    23.02 &    23.47 &    24.72  \\ 
LAE J221809.2+001242 &   3.096 &      60.0 &   4980.1 &   4972.2 &   4976.9 &    24.82 &    25.05 &    27.06  \\ 
LAE J221809.4+001358 &   3.105 &     124.5 &   4989.9 &      0.0 &   4989.9 &    24.96 &    25.49 &    26.77  \\ 
LAE J221811.0+002508 &   3.095 &     294.8 &   4978.9 &      0.0 &   4978.9 &    25.04 &    25.07 &    26.59  \\ 
LAE J221812.5+001433 &   3.095 &     164.1 &   4978.8 &      0.0 &   4983.9 &    23.57 &    24.15 &    25.25  \\ 
LAE J221813.2+002157 &   3.115 &     176.1 &   5003.6 &   4990.3 &   4999.2 &    25.22 &    25.32 &    26.37  \\ 
LAE J221813.8+000925 &   3.104 &     110.3 &   4989.8 &   4983.0 &   4987.9 &    24.83 &    24.83 &    27.93  \\ 
LAE J221813.9+002222 &   3.089 &       0.0 &   4971.3 &      0.0 &   4971.3 &    24.69 &     0.00 &     0.00  \\ 
LAE J221816.0+001412 &   3.086 &      74.2 &   4968.0 &   4957.4 &   4964.7 &    24.24 &    24.66 &    26.64  \\ 
LAE J221817.3+001209 &   3.087 &     330.9 &   4970.8 &      0.0 &   4966.6 &    24.45 &    24.78 &    25.77  \\ 
LAE J221818.2+001143 &   3.091 &     315.8 &   4975.6 &      0.0 &   4975.6 &    25.00 &    25.17 &    26.11  \\ 
LAE J221820.8+001257 &   3.109 &     119.2 &   4996.0 &   4986.8 &   4991.7 &    25.40 &    25.28 &    27.52  \\ 
LAE J221820.8+001241 &   3.103 &     342.0 &   4989.6 &   4978.0 &   4986.0 &    23.78 &    24.60 &    26.46  \\ 
LAE J221820.9+001031 &   3.097 &     178.3 &   4981.3 &      0.0 &   4981.3 &    24.99 &    25.50 &    28.16  \\ 

\tableline
\end{tabular}
\end{center}

(a) Redshift measured by the peak of the strongest emission-line component.
(b) FWHM of the strongest component after corrected for the instrumental profile. The zero value means that the line is not resolved.
(c) The central wavelength of the red peak in the multiple-component Gaussian fitting.
(d) The central wavelength of the blue peak in the multiple-component Gaussian fitting.
(e) The central wavelength of peak in the single-component Gaussian fitting.
(f) The aperture magnitude values are measured in the 2-arcsec diameter apertures.\end{table}

\tablenum{2}

%%%%%%%%% Table %%%%%%%%
\begin{table}
\begin{center}
\caption{Number of the LAEs with Characteristic Profile\label{tab2}}

\begin{tabular}{lccc}
\tableline\tableline
        &   All &  $NB497 < 25$ &   $NB497 < 25$, $BV-NB > 1.0$ \\ 
\tableline
Total              &    89 &  53           &     47 \\
"Red+Blue" profile &    39 &  28           &     26 \\
        
\tableline
\end{tabular}
\end{center}
\end{table}
%%%%%%%%% Table %%%%%%%%

\tablenum{3}

%%%%%%%%% Table %%%%%%%%
\begin{table}
\begin{center}
\caption{Objects Associated With Ly$\alpha$ Blobs\label{tab3}}

\begin{tabular}{lccc}
\tableline\tableline
  Name      &  FWHM (single Gaussian)  &  Ly$\alpha$ Blobs$^a$  & Profile$^b$\\ 
\tableline
            &  km s$^{-1}$   &     &  \\
\tableline

LAE J221658.4+002430  &  428.8       &   LAB34 &  No  \\
LAE J221706.7+002134  &  562.9       &   LAB27 &  Yes \\
LAE J221723.8+002155  &  474.3       &   LAB32 &  No  \\
LAE J221724.8+001717  &  474.4       &   LAB35 &  No  \\
LAE J221735.9+001559  &  696.0       &   LAB14 &  No  \\
LAE J221738.9+001102  &  641.2       &   LAB31 &  Yes \\
LAE J221740.9+001125  &  338.1       &   LAB7  &  No  \\
LAE J221759.2+002254  &  625.5       &   LAB28 &  Yes \\
LAE J221807.9+002317  &  582.4       &   LAB23 &  Yes \\
LAE J221808.3+001022  &  738.8       &   LAB15 &  Yes \\
LAE J221812.5+001433  &  222.1       &   LAB33 &  No  \\
LAE J221817.3+001209  &  661.5       &   LAB21 &  No  \\

\tableline
\end{tabular}
\end{center}
(a) Name of the associated Ly$\alpha$ Blobs listed in Matsuda et al. (2004)
(b) The column indicates whether the Ly$\alpha$ line profile shows the characteristic `strong red and weak blue' one (``Yes") or not (``No").

\end{table}
%%%%%%%%% Table %%%%%%%%

\end{document}